\documentstyle[din_a4,11pt,epsf,epsfig]{article}

\begin{document}

\pagestyle{empty}

\renewcommand{\thefootnote}{\fnsymbol{footnote}}

\setlength{\parindent}{0pt}
\vskip2cm
\hskip 10.5cm {\sl HUB-EP-96/30}
\vskip.0pt

\begin{center}

\vskip.5cm

{\Large \bf An Introduction to the Worldline Technique}\\
{\Large \bf for Quantum Field Theory Calculations
}
\\[1ex]

\vskip1.7cm

{\Large         Christian Schubert
\footnote{Supported by Deutsche Forschungsgemeinschaft.}
}\\[1.5ex]

\vskip.1cm 

{\it Humboldt Universit\"at zu Berlin\\
Invalidenstr. 110, D-10115 Berlin, Germany\\}
{schubert@qft2.physik.hu-berlin.de}

\vskip 2.9cm

{\sl\large 
Lectures given at the \\
{\sl XXXVI Cracow School of
Theoretical Physics,} \\
Zakopane, Poland, June 1 -- 10, 1996.
}

\vskip3.4cm
 
{\large\bf Abstract}

\end{center}

\begin{quotation}

\noindent
These two lectures give
a pedagogical introduction to
the ``string-inspired'' worldline
technique for perturbative calculations
in quantum field theory.  
This includes an overview over the
present range of its applications.
Several examples are calculated in
detail, up to the three-loop level.
The emphasis is on
photon scattering in quantum electrodynamics.

\end{quotation}

\clearpage

\renewcommand{\thefootnote}{\protect\arabic{footnote}}

\pagestyle{plain}

\setcounter{page}{1}
\setlength{\parindent}{0pt}

\newcommand{\be}{\begin{equation}}
\newcommand{\ee}{\end{equation}}
\newcommand{\bear}{\begin{eqnarray}}
\newcommand{\ear}{\end{eqnarray}}
\newcommand{\no}{\noindent}
\def\no{\noindent}
\def\non{\nonumber\\}
\def\sy{\scriptscriptstyle}
\def\dps{\displaystyle}
\def\D12{{\tau_1 - \tau_2}}
\def\D13{{\tau_1 - \tau_3}}
\def\D23{{\tau_2 - \tau_3}}
\def\half{{1\over 2}}
\def\third{{1\over3}}
\def\fourth{{1\over4}}
\def\fifth{{1\over5}}
\def\sixth{{1\over6}}
\def\seventh{{1\over7}}
\def\eigth{{1\over8}}
\def\ninth{{1\over9}}
\def\tenth{{1\over10}}
\def\pa{\partial}
\def\Gd{{\dot G_B}}
\def\g12{{\dot {G_B}_{12}}}
\def\g13{{\dot {G_B}_{13}}}
\def\g23{{\dot {G_B}_{23}}}
\def\Dab{{(x_a-x_b)}}
\def\Dsq{{({(x_a-x_b)}^2)}}
\def\lag{( -\partial^2 + V)}
\def\Tint{{\dps\int_{0}^{\infty}}
{dT\over T}e^{-m^2T}}
\def\Dx{\dps\int{\cal D}x}
\def\Dy{\dps\int{\cal D}y}
\def\Dpsi{\dps\int{\cal D}\psi}
\def\pint{{\dps\int}{dp_i\over {(2\pi)}^d}}
\def\Tr{{\rm Tr}\,}
\def\tr{{\rm tr}\,}
\def\e{\mbox{e}}
\def\g{\mbox{g}}
\def\O1{O($T^1$)}
\def\O2{O($T^2$)}
\def\O3{O($T^3$)}
\def\O4{O($T^4)}
\def\O5{O($T^5$)}
\def\Ge{\mbox{GeV}}
\def\dA{\partial^2}
\def\DA{\sqsubset\!\!\!\!\sqsupset}
\def\eins{  1\!{\rm l}  }
\def\freeexp{{\rm e}^{-\int_0^Td\tau {1\over 4}\dot x^2}}
\renewcommand{\thefootnote}{\fnsymbol{footnote}}
\newcommand{\GeV}{\mbox{GeV}}
\newcommand{\cL}{\cal L}
\newcommand{\Det}{\rm Det}
\newcommand{\PP}{\cal P}
\newcommand{\G}{{\cal G}}
\def\R{1\!\!{\rm R}}
\def\eins{  1\!{\rm l}  }
\newcommand{\symb}{\mbox{symb}}
\renewcommand{\arraystretch}{2.5}
\newcommand{\slD}{\raise.15ex\hbox{$/$}\kern-.57em\hbox{$D$}}

{\bf Introduction}
\vskip5pt
\no
Many one-loop amplitudes in
relativistic
quantum field theory admit representations in
terms of path integrals over the space
of closed loops 
~\cite{feynman50,fradkin66,friedbook,
kleinertbook,polbook,bardur,frgishbook}. 
Those path integrals
are closely related to the Feynman path integrals
used in nonrelativistic quantum mechanics, 
though they have
perhaps not reached quite the same practical
importance. Still there exists a considerable
amount of literature on applications of
particle path integrals to quantum field theory
(see e.g.
~\cite{halsie,hajese,alvarez-gaume,mckeon,kaktst,bpsn,rossch}), 
to which I cannot possibly do justice
here. 

The subject of these lectures rather is a
specific and novel way of evaluating this
type of path integral.
This method may be called
``string--inspired'' because it
is analogous to calculations in string
perturbation theory, and its development was
triggered by efforts 
~\cite{berkos,berntasi} to use the peculiar
organization of string amplitudes
to improve on the efficiency of calculations
in ordinary quantum field theory.
However this is to some extent accidental, and
the practical application of this technique
requires no knowledge of string theory.

The lectures are meant as an introduction for
practicioners in quantum field theory.
More effort will therefore be spent on
explaining the calculation rules of the worldline
formalism, than on rigorous derivations.

The first lecture is devoted to the theoretical
side. I will first sketch the
history of the subject, and
then give an overview over
the range of one-loop amplitudes
in field theory which are presently
accessible to this method. 
I concentrate then on the
explanation of an even more recent multiloop
generalization of this formalism
~\cite{schsch2,schsch3,dashsu,sato,rolsat,flscsc2,schsch4}.

The second lecture contains four 
detailed examples
of calculations performed with this
formalism, 
mainly taken from quantum electrodynamics:

\vspace{-5pt}
\begin{enumerate}
\item One -- loop QED vacuum polarization.
\vspace{-5pt}
\item One -- loop QED photon splitting.
\vspace{-5pt}
\item The two -- loop (scalar and spinor) QED
$\beta$ -- functions.
\vspace{-5pt}
\item The three -- loop scalar master integral.
\end{enumerate}
\no
\renewcommand{\baselinestretch}{1.0}

\vskip5pt
{\large\bf 1. First Lecture: Theory}
\vskip5pt
\setcounter{equation}{0}
\vskip5pt
\underline{\bf 1.1) The Bern-Kosower Formalism}
\vskip5pt
\no
It has been realized not too long
ago
that string theory may be used
as a guiding principle 
for finding new
and perhaps more
efficient parameter integral
representations for amplitudes in
ordinary quantum field theory.
The relevance of string theory for
this purpose resides in the well-known
fact that, in a sense which can be
made precise, string theory
contains ordinary field theory in
the limit where the tension of the string
becomes infinite.
In this limit, all massive
modes of the string get suppressed, and one
remains with the massless modes. Those
can be identified with ordinary massless particles
such as gauge bosons, gravitons, or massless
spin-$\half$ fermions. 

To use this fact for actual calculations
of amplitudes may appear an enormous detour. 
It is worthwhile nevertheless
due to the different organization of
string amplitudes, and of the 
different methods
used to calculate them. For details on string
perturbation theory, the reader may consult
~\cite{gsw}; let us mention here just
the following qualitative
points which will be important later on:

\vspace{-5pt}
\begin{enumerate}
\item Scattering amplitudes in string theory are usually
calculated in first, not second quantization.
\vspace{-5pt}
\item String theory calculations involve a much smaller
number of different ``Feynman diagrams''.
\end{enumerate}
\no
The second point is illustrated in fig. 1,
depicting a two-point 
``Feynman diagram'' for the
closed string. In the infinite string tension
limit this Riemann surface is squeezed to a graph,
though not to a single one --
two Feynman diagrams of different topologies
emerge.
\vspace{1cm}

\begin{figure}[h]
\begin{center}
\begin{picture}(0,0)%
\epsfig{file=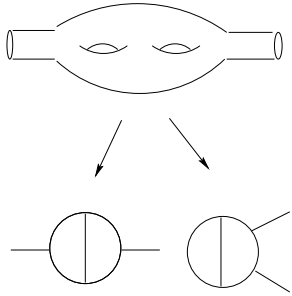}%
\end{picture}%
\setlength{\unitlength}{0.00087500in}%
\begingroup\makeatletter\ifx\SetFigFont\undefined
\def\x#1#2#3#4#5#6#7\relax{\def\x{#1#2#3#4#5#6}}%
\expandafter\x\fmtname xxxxxx\relax \def\y{splain}%
\ifx\x\y   
\gdef\SetFigFont#1#2#3{%
  \ifnum #1<17\tiny\else \ifnum #1<20\small\else
  \ifnum #1<24\normalsize\else \ifnum #1<29\large\else
  \ifnum #1<34\Large\else \ifnum #1<41\LARGE\else
     \huge\fi\fi\fi\fi\fi\fi
  \csname #3\endcsname}%
\else
\gdef\SetFigFont#1#2#3{\begingroup
  \count@#1\relax \ifnum 25<\count@\count@25\fi
  \def\x{\endgroup\@setsize\SetFigFont{#2pt}}%
  \expandafter\x
    \csname \romannumeral\the\count@ pt\expandafter\endcsname
    \csname @\romannumeral\the\count@ pt\endcsname
  \csname #3\endcsname}%
\fi
\fi\endgroup
\begin{picture}(3634,1536)(201,-850)
\end{picture}
\caption{\label{fig1} 
Infinite string tension limit of a string diagram}
\end{center}
\end{figure}
\vspace{-15pt}
\noindent
The first actual calculation done along these lines
was performed by
Metsayev and Tseytlin ~\cite{metseyt} in 1988,
who managed to
extract the correct one-loop $\beta$ -- function
coefficient for pure Yang-Mills theory from
the partition function of
an open string propagating in a Yang-Mills
background. This yielded also some explanation
for a fact which had been noted long before
~\cite{nepomechie83},
namely that this $\beta$ -- function,
when calculated in dimensional regularization, 
vanishes for $D=26$, the critical dimension of the
open string.

A systematic investigation of the infinite string tension
limit was undertaken in the following years 
by Bern and Kosower ~\cite{berkos}.
This was done again with a view on application
to QCD, however now to 
the calculation of complete on-shell scattering amplitudes.
Again the idea was to calculate, say,
gauge boson scattering amplitudes
in an appropriate (super) string model containing
SU(n) gauge theory, up to the point
where one has obtained
a parameter integral representation
for the amplitude considered. At this stage one 
performs the infinite string tension limit,
which should eliminite all contributions
due to propagating massive modes, and lead to
a parameter integral representation for the
corresponding field theory amplitude.

For the calculation of the string scattering amplitude,
one may use the Polyakov path integral.
In the simplest case, the closed
bosonic string
propagating in flat spacetime,  this is of
the form

\begin{equation}
\Bigl\langle
V_1\cdots V_N
\Bigr\rangle
\sim
\sum_{\rm top}
\int{\cal D}h
\int{\cal D}x(\sigma,\tau)
V_1\cdots V_N
\,
\e^{-S[x,h]}
\; .
\label{ppi}
\end{equation}
\no
Here the $\int{\cal D}x(\sigma,\tau)$
is over the space of all embeddings of the
string worldsheet with a fixed topology into
spacetime. The $\int{\cal D}h$ is over the
space of all worldsheet metrics, and the sum over
topologies 
$\sum_{\rm top}$ corresponds to the loop expansion in
field theory (fig. 2). 

\vspace{-25pt}
\begin{figure}[h]
\begin{center}
\begin{picture}(0,0)%
\epsfig{file=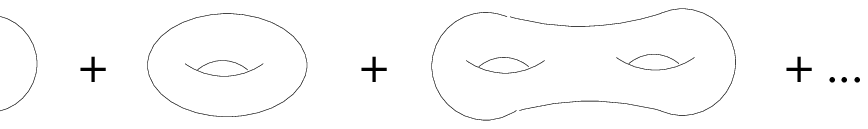}%
\end{picture}%
\setlength{\unitlength}{0.00087500in}%
\begingroup\makeatletter\ifx\SetFigFont\undefined
\def\x#1#2#3#4#5#6#7\relax{\def\x{#1#2#3#4#5#6}}%
\expandafter\x\fmtname xxxxxx\relax \def\y{splain}%
\ifx\x\y   
\gdef\SetFigFont#1#2#3{%
  \ifnum #1<17\tiny\else \ifnum #1<20\small\else
  \ifnum #1<24\normalsize\else \ifnum #1<29\large\else
  \ifnum #1<34\Large\else \ifnum #1<41\LARGE\else
     \huge\fi\fi\fi\fi\fi\fi
  \csname #3\endcsname}%
\else
\gdef\SetFigFont#1#2#3{\begingroup
  \count@#1\relax \ifnum 25<\count@\count@25\fi
  \def\x{\endgroup\@setsize\SetFigFont{#2pt}}%
  \expandafter\x
    \csname \romannumeral\the\count@ pt\expandafter\endcsname
    \csname @\romannumeral\the\count@ pt\endcsname
  \csname #3\endcsname}%
\fi
\fi\endgroup
\begin{picture}(3634,1536)(201,-850)
\end{picture}
\caption{\label{fig2} 
The loop expansion in string perturbation theory}
\end{center}
\end{figure}
\vspace{-15pt}

\noindent
\no
$S[x,h]$ is a free gaussian action,

\be
S[x,h]=-{1\over 4\pi \alpha'}
\int d\sigma d\tau
\sqrt{h}
h^{\alpha\beta}{\eta}_{\mu\nu}
\partial_{\alpha}x^{\mu}
\partial_{\beta}x^{\nu}
\; .
\label{S}
\ee\no
Here ${1\over 2\pi\alpha'}$ is the
string tension, and
the vertex operators 
$V_1,\ldots,V_N$ represent the scattering 
string states.
In the case of the open string, which is the more relevant one
for our purpose, the worldsheet has a boundary, and the vertex
operators are inserted on this boundary.
For instance, at the
one-loop level the worldsheet is just an annulus,
and a vertex operator may be integrated along either one of
the two boundary components.

\no
The vertex operators most relevant for us are
of the form

\bear
V_{\phi}&=&\int d\tau\,{\e}^{ik\cdot x(\tau)}
\label{scalarvertop}\\
V_A&=&\int d\tau \,T^{a}\varepsilon_{\mu}\dot x^{\mu}
{\e}^{ik\cdot x(\tau)}.
\label{gluonvertop}
\ear\no
They represent a scalar and a gauge boson with definite
momentum and polarization. $T^a$ is a generator
of the gauge group in some representation. 
The integration variable
$\tau$ parametrizes the boundary in question.
Since the action is gaussian, 
$\int{\cal D}x$ can be performed by Wick contractions,

\bear
\Bigl\langle
x^{\mu}(\tau_1)x^{\nu}(\tau_2)
\Bigr\rangle
=G(\tau_1,\tau_2)\eta^{\mu\nu}
\; ,
\label{wickstring}
\ear\no
where $G$ denotes the Green's function for the Laplacian
on the annulus, restricted to its boundary.

Performance of the one-loop path integral then
leads to the
following {\it Bern-Kosower Master Formula}
for the (single-trace partial contribution) to the
one-loop N-gluon scattering amplitude:

\begin{eqnarray}
A_{N;1}&\sim&{(\alpha')}^{{N\over2}-2}
Z
{\dps\int_{0}^{\infty}}{dT\over T}
{[4\pi T]}^{-{D\over 2}}
\prod_{i=1}^{N-1}d\tau_i\,\theta(\tau_i-\tau_{i+1})
\nonumber\\&&\!\!\!\!\!\!\!\!\times
\exp\biggl[\sum_{1\le i<j\le N}\alpha' G_{ij} p_ip_j
+\sqrt{\alpha'}\dot G_{ij}(p_i\varepsilon_j 
-p_j\varepsilon_i)
-\ddot G_{ij}\varepsilon_i\varepsilon_j\biggr]
\mid_{\rm multi-linear}.
\nonumber\\
\label{bkmaster}
\end{eqnarray}
\no 
Here it is understood that only the terms linear
in all the polarization vectors
$\varepsilon_1,\ldots,\varepsilon_N$
have to be kept. Dots generally denote a
derivative acting on the first variable,
$\dot G(\tau_1,\tau_2) = {\partial\over
{\partial\tau_1}}G(\tau_1,\tau_2)$, 
and we abbreviate
$G_{ij}:=G(\tau_i,\tau_j)$ etc.
$D$ is the spacetime dimension, and
Z denotes the partition function of the free string.
The $T$ -- integral is a remnant of the path integral over
the space of all metrics. 

This amplitude is still to be multiplied with
a colour trace factor ${\rm tr}(T^{a_1}\cdots
T^{a_N})$, and to be summed over non-cyclic permutations
(in the original Bern-Kosower approach colour
decomposition is used
for the adjoint representation, so that only the fundamental representation
matrices appear).

The main point  
to be noted is that this formula serves
as a unifying generating functional for all
one-loop N-point gluon amplitudes -- something for which no
known analogue exists in standard field theory.

The analysis of the infinite string tension limit,
for which I
refer the reader to 
~\cite{berkos,berntasi}, proceeds differently for the cases of
a spin-0, spin-$\half$, or spin-1 particle
circulating in the loop. 
The result of this analysis can be summarized in the
``Bern-Kosower Rules'', which allow one to construct
the final integral representations without referring
to string theory any more.
The relation of those parameter integrals to the ones
originating from the
corresponding Feynman diagram calculations has been
clarified in ~\cite{berdun}. Concerning the advantage of the
Bern-Kosower rules over the Feynman rules,
we mention the following points:

\vspace{-5pt}
\begin{enumerate}
\item Superior organization of gauge invariance.
\vspace{-5pt}
\item Absence of loop momentum, which reduces the number of
kinematic invariants from the beginning, and allows for a
particularly efficient use of the spinor helicity method.
\vspace{-5pt}
\item The method combines nicely with spacetime
supersymmetry.
\end{enumerate}
\vspace{-5pt}
\no
This has been demonstrated in both gluon ~\cite{bediko5glu}
and graviton ~\cite{bedush,dunnor} scattering calculations.
However, we are interested here mainly in a different property
of this formalism. To demonstrate this property, it suffices to
consider the simplest possible case of a Bern-Kosower
type formula, namely
the one-loop $N$ -- point amplitude for
$\phi^3$ -- theory. For this case, which was already
considered in 
~\cite{polbook}, one has
(up to a coupling
constant factor)

\begin{equation}
\Gamma^{(1)}(p_1,\ldots,p_N) =
{\dps\int_{0}^{\infty}}{dT\over T}
{[4\pi T]}^{-{D\over 2}}
\prod_{i=1}^N \int_0^T 
d\tau_i
\exp\biggl[\sum_{1\le i<j\le N}G_B(\tau_i,\tau_j) p_ip_j\biggr],
\nonumber\\
\label{scalarmaster}
\end{equation}
\no
where

\be
G_B(\tau_i,\tau_j)=\mid \tau_i-\tau_j\mid 
-{{(\tau_i-\tau_j)}^2\over T}
\, .
\label{defG}
\ee
\no
This parameter integral can, for any given fixed ordering
of the integration variables $\tau_{i_1}>\tau_{i_2}
>\cdots>\tau_{i_N}$, easily be identified with the
corresponding Feynman parameter integral
~\cite{berkos,vanholten-scalar} (the usual $\alpha_k$ -- parameters
just correspond to the differences $\tau_{i_k}-\tau_{i_{k+1}}$).
The complete integral therefore
does not represent any particular Feynman diagram, 
with a fixed ordering of the external legs, but
{\sl the sum of them} (fig. 3):

\begin{figure}[h]
\begin{center}
\begin{picture}(0,0)%
\epsfig{file=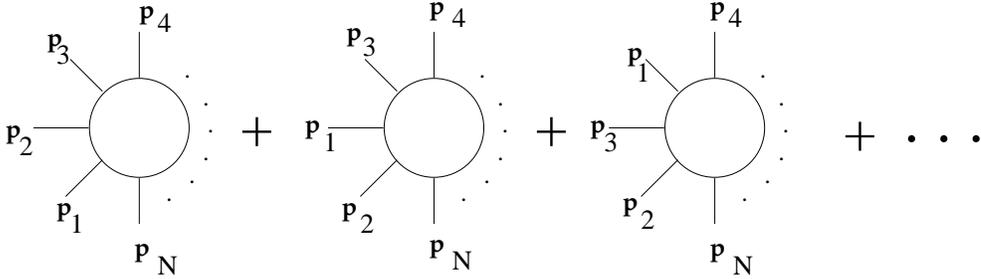}%
\end{picture}%
\setlength{\unitlength}{0.00087500in}%
\begingroup\makeatletter\ifx\SetFigFont\undefined
\def\x#1#2#3#4#5#6#7\relax{\def\x{#1#2#3#4#5#6}}%
\expandafter\x\fmtname xxxxxx\relax \def\y{splain}%
\ifx\x\y   
\gdef\SetFigFont#1#2#3{%
  \ifnum #1<17\tiny\else \ifnum #1<20\small\else
  \ifnum #1<24\normalsize\else \ifnum #1<29\large\else
  \ifnum #1<34\Large\else \ifnum #1<41\LARGE\else
     \huge\fi\fi\fi\fi\fi\fi
  \csname #3\endcsname}%
\else
\gdef\SetFigFont#1#2#3{\begingroup
  \count@#1\relax \ifnum 25<\count@\count@25\fi
  \def\x{\endgroup\@setsize\SetFigFont{#2pt}}%
  \expandafter\x
    \csname \romannumeral\the\count@ pt\expandafter\endcsname
    \csname @\romannumeral\the\count@ pt\endcsname
  \csname #3\endcsname}%
\fi
\fi\endgroup
\begin{picture}(3634,1536)(201,-850)
\end{picture}
\caption{\label{fig3} 
Sum of one-loop diagrams in $\phi^3$ -- theory.}
\end{center}
\end{figure}
\vspace{-10pt}

\noindent
This may not seem particularly relevant at the one-loop
level. However eq.(~\ref{scalarmaster})
holds off-shell, so that one can sew together, say, 
legs number $1$ and
$N$, and obtain the following sum of
two -- loop ($N-2$) -- point diagrams
(fig. 4):

\begin{figure}[h]
\begin{center}
\begin{picture}(0,0)%
\epsfig{file=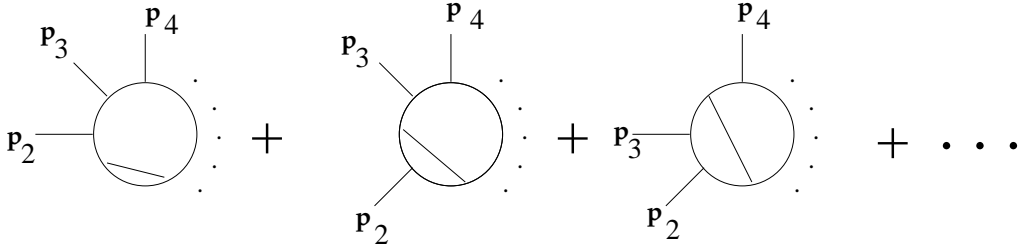}%
\end{picture}%
\setlength{\unitlength}{0.00087500in}%
\begingroup\makeatletter\ifx\SetFigFont\undefined
\def\x#1#2#3#4#5#6#7\relax{\def\x{#1#2#3#4#5#6}}%
\expandafter\x\fmtname xxxxxx\relax \def\y{splain}%
\ifx\x\y   
\gdef\SetFigFont#1#2#3{%
  \ifnum #1<17\tiny\else \ifnum #1<20\small\else
  \ifnum #1<24\normalsize\else \ifnum #1<29\large\else
  \ifnum #1<34\Large\else \ifnum #1<41\LARGE\else
     \huge\fi\fi\fi\fi\fi\fi
  \csname #3\endcsname}%
\else
\gdef\SetFigFont#1#2#3{\begingroup
  \count@#1\relax \ifnum 25<\count@\count@25\fi
  \def\x{\endgroup\@setsize\SetFigFont{#2pt}}%
  \expandafter\x
    \csname \romannumeral\the\count@ pt\expandafter\endcsname
    \csname @\romannumeral\the\count@ pt\endcsname
  \csname #3\endcsname}%
\fi
\fi\endgroup
\begin{picture}(3634,1536)(201,-850)
\end{picture}
\caption{\label{fig4} 
Sum of two -- loop diagrams with different topologies}
\end{center}
\end{figure}

\noindent
\no
Now this is obviously interesting, as we have at hand 
a single
integral formula for a sum containing diagrams of
different topologies. We may think of this as a remnant
of the ``less fragmented'' nature of string perturbation
theory mentioned before (fig. 1). 

Moreover, it calls certain well-known cancellations to mind
which happen in gauge theory due to the fact that the
Feynman diagram calculation splits a gauge invariant amplitude
into gauge non-invariant pieces. For instance, to obtain
the 3-loop $\beta$ -- function coefficient for
quenched QED, one needs to calculate the sum of diagrams
shown in fig. 5.

\begin{figure}[h]
\begin{center}
{}~\epsfig{file=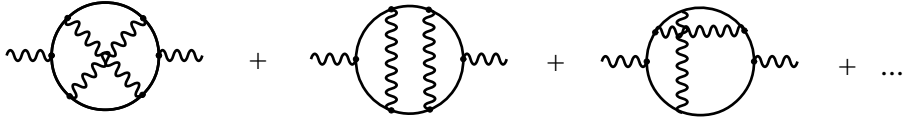}
\end{center}
\caption[]
{
\label{3-loopQEDbeta}
Sum of diagrams contributing to the 3-loop QED $\beta$ -- function.
}
\end{figure}

\no
Performing this calculation in, say, dimensional regularization, one finds
that

\begin{enumerate}
\item
All poles higher than ${1\over\varepsilon}$ cancel.

\item
Individual diagrams give contributions to 
the $\beta$ -- function proportional \break
\no$\phantom{ii)}$to $\zeta (3)$ 
which cancel in the sum.

\end{enumerate}
\no
The first property is known to hold to all orders of
perturbation theory ~\cite{jowiba}. 
Recently concepts from knot theory have been
used to establish a link between both
properties ~\cite{brdekr},
and thus to predict the rationality of the quenched
$\beta$ -- function to all orders.

It seems therefore very natural to apply the Bern-Kosower
formalism to this type of calculation. 
However, in its original version the
Bern-Kosower formalism was confined to tree--level
and one--loop amplitudes. A multiloop generalization
along the original lines has not yet been constructed,
though partial results exist ~\cite{roland92,dlmmr}. 
In the absence of better ideas
one could, of course, always start at the one-loop level,
and pursue the explicit sewing procedure indicated above.
This turns out to be unnecessarily clumsy, though; there
is a more efficient way of inserting propagators, which
will be explained in the third part of this lecture.
It is based on a purely field-theoretical approach to the
Bern-Kosower formalism, which was, at
the one-loop level, proposed by Strassler ~\cite{strassler92}.
This approach uses
a representation of one-loop amplitudes in terms
of first-quantized worldline path integrals, of a 
well-known type. I will first demonstrate this
method for the example of photon scattering in scalar
QED.

\no
The one-loop effective action for a Maxwell background
induced by a scalar loop may be written in terms of the
following 
first-quantized particle path integral 
(see e.g.~\cite{polbook,bpsn}):

\begin{equation}
\Gamma\lbrack A\rbrack = \int_0^{\infty}
{dT\over T} \, {\rm e}^{-m^2T}
\int {\cal D}x\, 
{\rm exp} \left[ 
- \int_0^T \!\!\! d\tau \left( {1\over 4}{\dot x}^2 
+ ieA_{\mu}\dot x^{\mu} 
\right) \right].
\label{scalarpi}
\end{equation}
\no
In this formula
$T$ is the usual Schwinger proper--time parameter,
and $m$ the mass of the particle circulating in the loop.
At fixed proper-time $T$, we have a
path integral over the space of 
periodic
functions $x^{\mu}(\tau )$ with period $T$,
describing all possible embeddings of the circle
with circumference $T$ into 
spacetime. 
We will generally use dimensional regularization, and
therefore work in $D$ dimensions from the beginning.
The spacetime metric is taken Euclidean.

The essence of the method is simple: We will evaluate
this path integral in a one-dimensional perturbation
theory. If we expand the
``interaction exponential'',

\be
{\rm exp}\Bigl[
-\int_0^Td\tau\, ieA_{\mu}\dot x^{\mu}
\Bigr]
=\sum_{N=0}^{\infty}
{{(-ie)}^N\over N!}
\prod_{i=1}^N
\int_0^Td\tau_i
\biggl[
\dot x^{\mu}(\tau_i)
A_{\mu}(x(\tau_i))
\biggr]
\label{expandint}
\ee\no
the individual terms correspond to Feynman diagrams
describing a fixed number of
interactions of the scalar loop with
the external field (fig. 6):

\begin{figure}[h]
\begin{center}
\begin{picture}(0,0)%
\epsfig{file=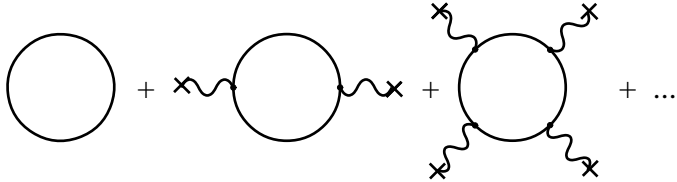}%
\end{picture}%
\setlength{\unitlength}{0.00087500in}%
\begingroup\makeatletter\ifx\SetFigFont\undefined
\def\x#1#2#3#4#5#6#7\relax{\def\x{#1#2#3#4#5#6}}%
\expandafter\x\fmtname xxxxxx\relax \def\y{splain}%
\ifx\x\y   
\gdef\SetFigFont#1#2#3{%
  \ifnum #1<17\tiny\else \ifnum #1<20\small\else
  \ifnum #1<24\normalsize\else \ifnum #1<29\large\else
  \ifnum #1<34\Large\else \ifnum #1<41\LARGE\else
     \huge\fi\fi\fi\fi\fi\fi
  \csname #3\endcsname}%
\else
\gdef\SetFigFont#1#2#3{\begingroup
  \count@#1\relax \ifnum 25<\count@\count@25\fi
  \def\x{\endgroup\@setsize\SetFigFont{#2pt}}%
  \expandafter\x
    \csname \romannumeral\the\count@ pt\expandafter\endcsname
    \csname @\romannumeral\the\count@ pt\endcsname
  \csname #3\endcsname}%
\fi
\fi\endgroup
\begin{picture}(3634,1536)(20,-850)
\end{picture}
\caption{\label{fig6} 
Expanding the path integral in powers of the
background field}
\end{center}
\end{figure}
\vspace{-5pt}

\no
The corresponding $N$ -- photon
scattering amplitude is then obtained by
specializing to a background
consisting of 
a sum of plane waves with definite
polarizations,

\be
A_{\mu}(x)=
\sum_{i=1}^N
\varepsilon_{\mu}^{(i)}
\e^{ik_i\cdot x}
, 
\label{planewavebackground}
\ee\no
and picking out the term containing every
$\varepsilon^{(i)}$ once (this also removes
the ${1\over N!}$ in eq.(~\ref{expandint})).
We find thus 
exactly the same 
photon
vertex operator used in string perturbation
theory, eqs.(\ref{scalarvertop}),(\ref{gluonvertop}), inserted on a circle
instead on the boundary of the annulus.
But 
in the infinite string tension limit
the annulus gets squeezed to a circle; 
we may therefore think of the path integral
eq.(\ref{scalarpi}) as the infinite string
tension limit of the corresponding
Polyakov path integral described above.
Again the evaluation of the path integral at fixed values of
the parameters reduces to Wick-contracting

\be
\biggl\langle
\dot x_1^{\mu_1}\e^{ik_1\cdot x_1}
\cdots
\dot x_N^{\mu_N}\e^{ik_N\cdot x_N}
\biggr\rangle
\label{scalqedwick}
\ee
\no
The Green's function to be used is now
simply the one for the second-derivative
operator, acting on
periodic functions. To derive 
this Green's function, first observe that 
$\int{\cal D}x(\tau)$ contains the constant functions,
which we must get rid of to obtain a well-defined
inverse. Let us therefore restrict our integral
over the space of all loops by fixing the average
position $x_0^{\mu}$ of the loop:

\begin{eqnarray}
{\displaystyle\int}{\cal D}x &=&
{\displaystyle\int}d x_0{\displaystyle\int}{\cal D} y,
\label{split}\\
x^{\mu}(\tau) &=& x^{\mu}_0  +
y^{\mu} (\tau ),\\
\int_0^T d\tau\,   y^{\mu} (\tau ) &=& 0\quad .\label{constraint}
\end{eqnarray}
\noindent
For effective action calculations this reduces the
effective action to the effective Lagrangian,

\vspace{-18pt}
\begin{equation}
\Gamma=\int d^Dx_0\,{\cal L}(x_0).
\label{EAtoEL}
\end{equation}
\no
In scattering amplitude calculations,
the integral over $x_0$ just gives
momentum conservation,

\vspace{-18pt}
\be
\int d^D x_0\,
\e^{ix_0\cdot \sum_{i=1}^N k_i}
= {(2\pi)}^D\delta\Bigl(\sum_{i=1}^N k_i\Bigr)
\label{momentumcons}
\ee\no
The reduced path integral $\int{\cal D}y(\tau)$
has an invertible kinetic operator. This
inverse is easily constructed using the
eigenfunctions of the derivative operator on the circle
with circumference $T$, $\lbrace
\e^{2\pi in{\tau\over T}},n\in {\bf Z}\rbrace$, and leads to

\vspace{-8pt}
\be
2\bigl\langle\tau_1\mid
{\Bigl({d\over d\tau}\Bigr)}^{-2}
\mid\tau_2\bigr\rangle
= 2T
\sum_{{n=-\infty}\atop{n\ne 0}}^{\infty}
{1\over {(2\pi in)}^2}
\e^{2\pi in{{\tau_1-\tau_2}\over T}}
=
G_B(\tau_1,\tau_2)-{T\over 6}.
\label{calcG}
\ee
\vspace{-8pt}

\no
Here $G_B$ is the function which was introduced already
in eq.(\ref{defG}). 
The constant $-T/6$ drops out
of all calculations once momentum conservation is applied, and
is therefore usually deleted at the beginning.

Working out eq.(\ref{scalqedwick}) by Wick-contractions using the
function $G_B$ leads to a parameter integral representation
for the 
one-loop N-photon scattering amplitude 
identical with the one encoded in the Bern-Kosower
master formula for this special case. We will look at this
parameter integral
in more detail later on.

Alternatively, one can also apply this method for calculating
the higher derivative expansion of the effective action
$\Gamma(A)$ ~\cite{schsch1,flscsc1,fhss}. This can be done
in a manifestly gauge invariant way using Fock-Schwinger gauge
centered at $x_0$.

\vskip15pt
\underline{\bf 1.2) One-Loop Wordline Path Integrals and Correlators}
\vskip5pt

The representation of amplitudes in quantum field theory
in terms of first-quantized particle path integrals is an old
and well-studied subject, and I cannot possibly survey 
the literature here. Let me just mention a few
contributions
which were particularly
important for the development of the subject:

\vspace{-3pt}
\begin{enumerate}
\vspace{-5pt}
\item
R.P. Feynman 1950 \cite{feynman50} (path integral representation of the
scalar propagator).
\vspace{-5pt}
\item
E.S. Fradkin 1967 \cite{fradkin66} (path integral representation of the
electron propagator).
\vspace{-5pt}
\item
Berezin and Marinov 1977 \cite{bermar}, L. Brink, P. Di Vecchia
and  P. Howe 1977 \cite{brdiho} (superparticle Lagrangians).
\vspace{-5pt}
\item
L. Alvarez-Gaum\'e 1983 \cite{alvarez-gaume} (application 
of superparticle path integrals to anomaly calculations).
\end{enumerate} 
\vspace{-5pt}
\no
The following overview is restricted to those cases
which are presently relevant for the ``string-inspired''
technique.

\vfill\eject
\no\underline{\it The Free Gaussian Path Integral}
\vskip.3cm

With our conventions, the free coordinate path integral is

\begin{equation}
\int {\cal D}y\, {\rm exp} \left[ 
- \int_0^T \left( {{\dot y}^2\over 4} \right) \right]
= {\Bigl(4\pi T\Bigr)}^{-{D\over 2}}
\label{freepi}
\end{equation}
\no
All other free path integrals are normalized to unity.

\vspace{10pt}
\no\underline
{\it Scalar Field Theory with a self-interaction potential $U(\phi)$}
\vskip.3cm

For a (real) scalar field
with a self-interaction $U(\phi)$,
the path integral representation of the one-loop
effective action analogous to
eq.(~\ref{scalarpi}) reads

\begin{equation}
\Gamma[\phi] = {1\over 2} \int_0^{\infty}{dT\over T}\, 
{\rm e}^{-m^2T} \int {\cal D}x\, {\rm exp} \left[ 
-\int_0^T \left( {{\dot x}^2\over 4} 
+ U''\left( \phi (x) \right)\right) \right]
\label{scalareffact}
\end{equation}
\no
For example, for $U(\phi)= {\lambda\over 3!}\phi^3$
one has an interaction exponential

$$\exp\Bigl[
-\lambda\int_0^Td\tau\phi(x(\tau))
\Bigr],
$$
\no
leading to a scalar vertex operator $V_{\phi}$ 
such as in
eq.(~\ref{scalarvertop}). Wick-contraction
of $N$ such operators, 
using

\begin{equation}
\langle y^\mu(\tau_1) \, y^\nu(\tau_2)\rangle
 = -g^{\mu\nu} \, G_B(\tau_1,\tau_2)
\label{wickscalar}
\end{equation}
\no
then leads to eq.(~\ref{scalarmaster}).

\vskip.2cm
\no\underline
{\it Photon Scattering in Quantum Electrodynamics}
\vskip.3cm

We have already discussed 
photon scattering in scalar QED. 
Eq.(~\ref{scalarpi}) generalizes to the
Dirac fermion loop as follows (see e.g. ~\cite{polbook}):

\begin{equation}
\Gamma\lbrack A\rbrack =\! -2\! \int_0^{\infty}
{dT\over T} {\rm e}^{-m^2T}
\int {\cal D}x\,{\cal D}\psi\,
{\rm exp}\left[ -\int_0^T \!\!\!d\tau
\left( {1\over 4}{\dot x}^2\! +\!{1\over2}\psi\dot\psi
\!+\!ieA_{\mu}\dot x^{\mu}\!-\!ie\psi^{\mu}F_{\mu\nu}\psi^{\nu}
\right)\right] \nonumber\\
\label{spinorpi}
\end{equation}
\no
The ``bosonic'' coordinate path integral ${\cal D}y$ is identical with
the scalar QED one. The additional 
``fermionic'' ${\cal D}\psi$ -- integration
is over anticommuting Grassmann -- valued functions, 
obeying
antiperiodic boundary conditions, 
$\psi^{\mu}(T)=-\psi^{\mu}
(0)$. We may think of this double path integral as breaking up a
Dirac spinor into a ``convective part'' and a ``spin part''.
The global factor of $(-2)$ compared
to the complex scalar loop case is due to the different statistics and
degrees of freedom. 
The photon vertex operator acquires an additional Grassmann piece,

\be
V_{A}=
\int_0^T d\tau\,\Bigl(\varepsilon_{\mu}\dot x^{\mu}
+2i\varepsilon_{\mu}\psi^{\mu}
k_{\nu}\psi^{\nu}\Bigr)
{\rm e}^{ik\cdot x(\tau)}
\label{photonvertopspin}
\ee\no
The Grassmann path integral is performed using the
Wick contraction rule

\begin{eqnarray}
\langle \psi^\mu(\tau_1)\, \psi^\nu(\tau_2)\rangle
&=& {1\over 2}g^{\mu\nu} \, G_F(\tau_1,\tau_2),
\label{wickgrassmann}\\
G_F(\tau_1,\tau_2)&=&{\rm sign}(\tau_1,\tau_2). 
\label{defGF}
\end{eqnarray}
\no
Before Wick--contracting two Grassmann fields, they must be
made adjacent using the anticommutation property.
The worldline action has a ``worldline
supersymmetry'', namely

\vspace{-10pt}
\be
\delta_{\eta}x^{\mu} = - 2\eta\psi^{\mu}, \,\,\,\,\,
\delta_{\eta}\psi^{\mu} = \eta \dot x^{\mu},
\label{susytrafo}
\ee
\no
with some Grassmann constant $\eta$.
It makes therefore sense to combine
$y$ and $\psi$ into a superfield.
Defining

\vspace{-5pt}
\begin{equation}
X^{\mu} = x^{\mu}\!+\!\sqrt 2\theta\psi^{\mu}=x_0^{\mu}\!+\!Y^{\mu},
\,\,\,\,D={\partial\over{\partial\theta}}\!
-\!\theta{\partial\over{\partial\tau}},\,\,\,\,\,
\int d\theta\,\theta = 1\, ,
\label{defsuper}
\end{equation}
\no
eq.(~\ref{spinorpi}) can be rewritten as a super path integral,

\vspace{-5pt}
\begin{equation}
\Gamma\lbrack A\rbrack = 
\! -2\! \int_0^{\infty} {dT\over T} {\rm e}^{-m^2T}
\int {\cal D}X {\rm exp} \left\{ - 
\int_0^T \!\!\!\!d\tau\int\!\!d\theta \, \left[\!-
{1\over 4}X\,D^3X 
\!-\!ie DX^{\mu}A_{\mu}(X) \right] \right\} 
\label{spinorsuperpi}
\end{equation}
\no
The two Wick contraction rules are then
combined into a single rule for the field $Y$,

\vspace{-5pt}
\bear
\langle Y^{\mu}(\tau_1,\theta_1)\, Y^{\nu}(\tau_2,\theta_2)\rangle
&=&- g^{\mu\nu}\, \hat G(\tau_1,\theta_1;\tau_2,\theta_2),
\label{superwick}\\
\hat G(\tau_1,\theta_1;\tau_2,\theta_2)
&=&
G_B(\tau_1,\tau_2) +\theta_1\theta_2
G_F(\tau_1,\tau_2).
\label{superprop}
\ear
\no
The photon vertex operator eq.(~\ref{photonvertopspin})
for the spinor loop is rewritten as

\be
V_A=
-\int_0^T d\tau d\theta \,\varepsilon_{\mu}
DX^{\mu}{\rm exp}[ikX].
\label{photonvertopsup}
\ee\no
This superfield reformulation makes no difference for the final
integral representations obtained, but is 
often useful for
keeping intermediate expressions compact.

\vskip.2cm
\no\underline
{\it Photon Scattering in a Constant External Field ${\bar F}$}
\vskip.3cm

An additional
background field $\bar A^{\mu}(x)$, 
with constant field strength tensor
$\bar F_{\mu\nu}$, can be completely
taken into account by changing the
worldline Green's functions, and
the path integral determinant
~\cite{cadhdu,shaisultanov,rescsc,adlsch}.
The Green's functions become
~\cite{rescsc}
\vspace{-10pt}
\begin{eqnarray}
{\cal G}_{B}(\tau_1,\tau_2) &=&
{1\over 2{(e\bar F)}^2}
\biggl({e\bar F\over{{\rm sin}(e\bar FT)}}
{\rm e}^{-ie\bar FT\dot G_{B12}}
+ie\bar F\dot G_{B12} -{1\over T}\biggr)
\non
\dot{\cal G}_B(\tau_1,\tau_2)
&=&
{i\over e\bar F}
\biggl({e\bar F\over{{\rm sin}(e{\bar F}T)}}
{\rm e}^{-ie\bar FT\dot G_{B12}}-{1\over T}\biggr)
\non
\ddot{\cal G}_{B}(\tau_1,\tau_2)
&=& 2\delta_{12} -2{e\bar F\over{{\rm sin}(e\bar FT)}}
{\rm e}^{-ie\bar FT\dot G_{B12}}
\non
{\cal G}_{F}(\tau_1,\tau_2) &=&
G_{F12}
{{\rm e}^{-ie\bar FT\dot G_{B12}}\over {\rm cos}(e\bar FT)}
\non
\label{calGBGF}
\end{eqnarray}
\noindent
These expressions should be understood as power
series in the field strength matrix $\bar F$
(note that eqs.(~\ref{calGBGF})
do {\sl not} assume 
invertibility of $\bar F$).
The background field breaks the Lorentz invariance, so
that the
generalized Green's functions are nontrivial
Lorentz matrices in general. Accordingly,
the Wick contraction rules 
eqs.(\ref{wickscalar}),(\ref{wickgrassmann})
have to be rewritten as

\begin{eqnarray}
\langle y^{\mu}(\tau_1)y^{\nu}(\tau_2)\rangle
&=&
-{\cal G}_B^{\mu\nu}(\tau_1,\tau_2),\nonumber\\
\langle\psi^{\mu}(\tau_1)\psi^{\nu}(\tau_2)\rangle
&=&
\frac{1}{2}{\cal G}_F^{\mu\nu}(\tau_1,\tau_2).
\label{exfieldGreen's}
\end{eqnarray}
\noindent
Again momentum conservation leads to the freedom
to subtract from
${\cal G}_B$ its
constant coincidence limit,

\begin{equation}
{\cal G}_{B}(\tau,\tau)=
{1\over 2{(e\bar F)}^2}
\biggl(e\bar F\cot(e\bar FT)-{1\over T}
\biggr)
\label{coincidence1}
\end{equation}
\noindent
To correctly obtain this and other coincidence
limits, one has to set 

$${\rm sign}(\tau,\tau) =0,$$ 

\no
and adopt
the general rule that coincidence limits must always be
taken $\sl after$ derivatives. It follows that, for instance,

\be
\dot G_B(\tau,\tau)=0, \,\,\,\,\,\,
\dot G_B^2(\tau,\tau)=1.
\label{coincidencerules}
\ee
\noindent
The change of the path integral determinant 
eq.(~\ref{freepi}) induced
by the external field is

%

\bear
\!\!\!\!
{(4\pi T)} ^{-{D\over 2}}
&\rightarrow&
{(4\pi T)}^{-{D\over 2}}
{\rm det}^{-{1\over 2}}
\biggl[{\sin(e\bar FT)\over {e\bar FT}}
\biggr] \quad\qquad{\rm (Scalar\; QED)}
\label{scaldetext}\\
\!\!\!\!
{(4\pi T)}^{-{D\over 2}}
&\rightarrow&
{(4\pi T)}^{-{D\over 2}}
{\rm det}^{-{1\over 2}}
\biggl[{\tan(e\bar FT)\over {e\bar FT }}
\biggr] \quad\qquad{\rm (Spinor\; QED)}
\label{spindetext}
\ear\no

\vskip.2cm

\no\underline
{\it Scalar or Spinor Loop Contribution to Gluon Scattering}
\vskip.3cm

\vskip.2cm
The path integrals for the external gluon case differ from 
eqs.(~\ref{scalarpi}),(~\ref{spinorpi})
only by path-ordering of the exponentials, and the
addition of a global colour trace. 
For a given $N$--gluon amplitude, this trace 
factors out as ${\rm tr}(T^{a_1}\cdots T^{a_N})$, where the
$T^{a_i}$ are the
colour matrices carried by the gluon vertex operators,
eq.(~\ref{gluonvertop}).
The path--ordering leads to ordered $\tau$ -- integrals
$\int\prod_{i=1}^{N-1}d\tau_i\theta(\tau_i-\tau_{i+1})$
such as in eq.(~\ref{bkmaster}) (translation invariance in
$\tau$ is used for setting $\tau_N=0$). 

In the fermion loop case, the worldline Lagrangian
eq.(~\ref{spinorpi}) now contains a term
$\psi^{\mu}[A_{\mu},A_{\nu}]\psi^{\nu}$
which,
in the component formalism,
forces one to introduce an additional
two -- gluon vertex operator 
besides eq.(~\ref{photonvertopspin})
~\cite{strassler92}. 
This is not necessary in the superfield formalism, 
where the super vertex operator 
eq.(~\ref{photonvertopsup})
remains sufficient. 
Here the only change is that a 
suitable supersymmetric
generalization
of the above $\theta$ -- functions is needed.
This is

\be
\theta(\hat\tau_{ij})=
\theta(\tau_i-\tau_j)+\theta_i\theta_j
\delta(\tau_i-\tau_j),
\label{supertheta}
\ee\no
where $\hat\tau_{ij}\equiv \tau_i-\tau_j+\theta_i\theta_j$.
The
nonabelian commutator terms above 
are then generated by the $\delta$ -- function terms
~\cite{andtse}.

\vskip.5cm
\no\underline
{\it Gluon Loop Contribution to Gluon Scattering}
\vskip.3cm
The treatment of the 
gluon loop case in this formalism 
is considerably more involved than the
scalar and spinor loop cases. 
This is due to the fact that the consistent
coupling of a spin-1 path integral
to a spin-1 background requires 
the introduction of auxiliary degrees of freedom,
whose contributions have to be removed later on.
I will just write down the appropriate path integral,
and refer the reader to
~\cite{strassler92,rescsc}
both for its derivation and evaluation. It reads

\bear
\Gamma[A]&=&\frac{1}{2}\lim_{C\to\infty}
\int^\infty_0\frac{dT}{T}\exp\Bigl[-CT(\frac{D}{2}-1)\Bigr]
\int_P{\cal D}x\,
\frac{1}{2}\Bigl(\int_A-\int_P\Bigr)
{{\cal D}}\psi{\cal D}\bar\psi
\nonumber\\
&&\!\!\!\!\!\!\!\!\!\!\times 
{{\tr}}{\PP}
\exp\biggl\lbrace -\int^T_0
d\tau\biggl[\frac{1}{4}\dot x^2+ig A_\mu\dot x^\mu
+\bar\psi^\mu\Bigl[(\partial_\tau-C)
\delta_{\mu\nu}-2ig F_{\mu\nu}\Bigr]
\psi^\nu\biggr]
\biggr\rbrace .\non
\label{gluonpi}
\ear
\no
The Grassmann path integral now appears both with
antiperiodic (``A'') and periodic (``P'') boundary
conditions. $\cal P$ denotes the path-ordering operator.
This path integral actually describes a whole
multiplet of $p$ -- forms, $p=0,\ldots,D$ circulating in
the loop; the role of the limit $C\rightarrow\infty$
is to suppress all contributions from $p\ge 2$, and the 
contributions from the zero form cancel out in the combination
$\int_A-\int_P$.

\vskip.2cm
\no\underline
{\it Pseudoscalar Yukawa Coupling}
\vskip.3cm

We consider now a pseudoscalar 
$\phi$ interacting with a Dirac fermion $\psi$
via the $\lambda\bar\psi\phi\gamma_5\psi$ -- vertex. 
In contrast to the
gauge path integral, the following path integral representation
for the
one-loop effective action induced for the pseudoscalar field
by a spinor loop was found only recently ~\cite{mnss1}:

\begin{eqnarray}
\Gamma(\phi) &=&\!-2\!\int_0^{\infty} {dT\over T}
\int{\cal D}x\,{\cal D}\psi\,{\cal D}\psi_5\,
{\rm e}^{-S_{\rm Yps}[x,\psi,\psi_5]}\nonumber\\
S_{\rm Yps}[x,\psi,\psi_5] &=& \!
\int_0^T d\tau\left\{
\frac{\dot x^2}{4}
\!+\!{1\over 2}\psi\dot\psi
\!+\!{1\over 2}\psi_5\dot\psi_5
\!+\!m^2\!+\!\lambda^2 \phi^2(x)
\!+\!2i \lambda  \psi_5 \psi \cdot \partial \phi(x)\right\}
\non
\label{mass}
\end{eqnarray}
\no
Besides $y$ and $\psi$
there is now a second fermionic field $\psi_5$.
It is a Lorentz scalar, and 
has the same correlator as $\psi$:

\begin{equation}
\langle \psi_5(\tau_1) \, \psi_5(\tau_2) \rangle
= \frac{1}{2} \, G_F(\tau_1,\tau_2)
\label{correlationfn}\nonumber
\end{equation}

\vfill\eject
\no\underline
{\it Scalar Yukawa Coupling}
\vskip.3cm
\noindent
The case of the scalar -- fermion coupling
$\lambda\bar\psi\phi\psi$ differs from the pseudoscalar
case only by the following change of the action,

\begin{equation}
S_{Yps}\rightarrow S_{Ys}=S_{Yps}+2\lambda m\phi(x)
\label{SSY}
\end{equation}
\no
For the combination of both couplings see ~\cite{mnss1}, for
the nonabelian case ~\cite{dhogag1,vanholten}.

To conclude this section, let us mention that the representation
of axial vector couplings on the worldline has been studied
in ~\cite{mnss2, dhogag2,gagne}. A treatment of 
external fermions
along the present lines is still lacking.

\vskip20pt
\underline{\bf 1.3) Multiloop Generalization}
\vskip8pt

\no
If one wishes to generalize Strassler's 
approach beyond one loop,
obviously one needs to know how the one-loop
Green's function $G_B$, defined on the circle,
generalizes to
higher order graphs.
As a first step in the construction of such
``multiloop worldline Green's functions''
~\cite{schsch2},
we will ask the following simple question:
How does the Green's function 
$G_B(\tau_1,\tau_2)$ 
between two fixed points $\tau_1,\tau_2$ on
the circle change, if we insert, between
two other points $\tau_a$ and $\tau_b$,
a (scalar) propagator of fixed proper-time length
$\bar T$ (fig. 7)? 

\par
\begin{figure}[ht]
\begin{center}
\vskip-6.cm
\epsffile{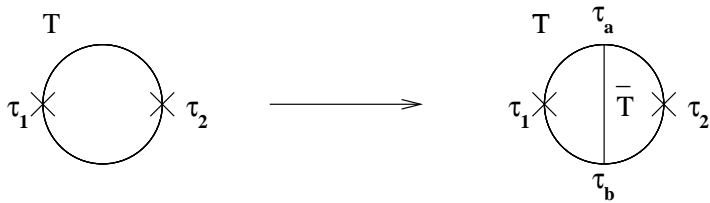}
\end{center}
\caption[dum]{Change of the one-loop Green's function
by a propagator insertion
\hphantom{xxxxxxxxxxxxxxx}}
\label{fig7}
\end{figure}
\par
\vspace{-6pt}

\no
To answer this question, let us start with the
worldline-path integral representation for the
one-loop two-point -- amplitude in 
$\phi^3$ -- theory,
and sew together the two external
legs. The result is, of course,
the vacuum path
integral with a propagator insertion:

\begin{equation}
\Gamma^{(2)}_{\rm vac} =
\int_0^{\infty}{{dT}\over T}{\rm e}^{-m^2T}
{\dps\int}
{\cal D} x
\int_0^T d\tau_a \int_0^T d\tau_b
\,\langle \phi(x(\tau_a))\phi(x(\tau_b))
\rangle
\,\exp\biggl[- \int_0^T d\tau
{{\dot x}^2\over 4}
\biggr]
\label{pi+prop}
\end{equation}
\no
Here
$\langle \phi(x(\tau_a))\phi(x(\tau_b))
\rangle$
is the x--space scalar propagator in $D$
dimensions, which, if we specialize to the
massless case for a moment, would read

\begin{equation}
\langle \phi(x(\tau_a))\phi(x(\tau_b))
\rangle = {{\Gamma ({D\over 2}-1)}\over
{4{\pi}^{D\over 2}
{\Bigl[{(x_a-x_b)}^2\Bigr]}^{({D\over 2} -1)}}}
\label{masslessscalprop}
\end{equation}
\no
Clearly
this form of the propagator is not suitable for
calculations in our auxiliary
one-dimensional field theory.
We will therefore again make use of the
Schwinger proper-time representation,

\begin{equation}
\langle\phi(x(\tau_a))\phi(x(\tau_b))\rangle
=\int_0^{\infty} d\bar T 
e^{-{\bar m}^2\bar T}
{(4\pi \bar T)}^{-{D\over 2}}
{\rm exp}\Biggl
[-{{(x(\tau_a)-x(\tau_b))}^2\over 4\bar T}
\Biggr ]\, ,
\label{intprop}
\end{equation}
\no
where the mass was also reinstated.
We have then the following path integral
representation for the two -- loop
vacuum amplitude:

\begin{eqnarray}
\Gamma_{\rm vac}^{(2)}&=&
\int_0^{\infty}{{dT}\over T}{\rm e}^{-m^2T}
\int_0^{\infty} d\bar T{(4\pi \bar T)}^{-{D\over 2}} 
e^{-{\bar m}^2\bar T}
\int_0^T d\tau_a \int_0^T d\tau_b
\nonumber\\
&&\times
{\dps\int}
{\cal D} x
\,\exp\Biggl [- \int_0^T d\tau\,
{{\dot x}^2\over 4}-
{{(x(\tau_a)-x(\tau_b))}^2\over 4\bar T}
\Biggr ]\nonumber\\
\label{pi+B}
\end{eqnarray}
\no
The propagator insertion has, for fixed parameters
$\bar T,\tau_a,\tau_b$, just produced an additional
contribution to the original free worldline action. 
Moreover, this term is quadratic 
in $x$, so we may hope to absorb it into the
free worldline Green's function. For this
purpose, it is useful to introduce
an integral operator $B_{ab}$
with integral kernel

\begin{equation}
B_{ab}(\tau_1,\tau_2) = 
\Bigl[\delta(\tau_1 - \tau_a)-\delta(\tau_1-\tau_b)\Bigr]
\Bigl[\delta(\tau_a - \tau_2)-\delta(\tau_b-\tau_2)\Bigr]
\quad 
\label{defB}
\end{equation}
\no
($B_{ab}$ acts trivially on Lorentz indices).
We may then rewrite

\begin{equation}
{(x(\tau_a)-x(\tau_b))}^2
= 
\int_0^T d\tau_1 \int_0^T d\tau_2\,
x(\tau_1)B_{ab}(\tau_1,\tau_2)x(\tau_2)
\quad .
\label{useB}
\end{equation}
\no
Obviously, the presence of the
additional term corresponds
to changing the defining equation for $G_B$,
eq.(~\ref{calcG}),
to
\begin{equation}
G_B^{(1)}(\tau_1,\tau_2)=2
\bigl\langle\tau_1\mid
{\Bigl({d^2\over {d\tau}^2}-{B_{ab}\over \bar T}\Bigl)}^{-1}
\mid\tau_2\bigr\rangle .
\label{defG(1)}
\end{equation}
\no
After eliminating the zero-mode
as before,
this modified equation can be solved simply by
summing a geometric series:

\begin{equation}
{\Bigl({d^2\over {d\tau}^2}-{B_{ab}\over \bar T}\Bigl)}^{-1}
= {\Bigl({d\over {d\tau}}\Bigr)}^{-2}
+
{\Bigl({d\over {d\tau}}\Bigr)}^{-2}{B_{ab}\over \bar T}
{\Bigl({d\over {d\tau}}\Bigr)}^{-2}
+
{\Bigl({d\over {d\tau}}\Bigr)}^{-2}{B_{ab}\over \bar T}
{\Bigl({d\over {d\tau}}\Bigr)}^{-2}{B_{ab}\over \bar T}
{\Bigl({d\over {d\tau}}\Bigr)}^{-2}
+\cdots,
\label{sumseries}
\end{equation}
\no
leading to ~\cite{schsch2}

\begin{equation}
G_B^{(1)}(\tau_1,\tau_2)=
G_B(\tau_1,\tau_2) + \half
{{[G_B(\tau_1,\tau_a)-G_B(\tau_1,\tau_b)]
[G_B(\tau_a,\tau_2)-G_B(\tau_b,\tau_2)]}
\over
{{\bar T} + G_B(\tau_a,\tau_b)}}
\label{G(1)} 
\end{equation}
\no

The worldline Green's function between points
$\tau_1$ and $\tau_2$ is 
thus simply the one-loop Green's function plus  
one additional piece, which takes the effect
of the insertion
into account. Observe that this piece
can still be written in terms of the various
one-loop Green's functions $G_{Bij}$. 
However it is not a function of
$\tau_1-\tau_2$ any more, nor is its
coincidence limit a constant.

Knowledge of this Green's function is not quite enough
for performing two-loop calculations. We also need to know
how the path integral determinant is changed by
the propagator insertion. 
Using the $ln\,det=tr\,ln$ -- formula, 
this can again be calculated easily, and yields

\begin{equation}
{{
{\dps\int}
{\cal D} y
\,\exp\Bigl [- \int_0^T d\tau
{{\dot y}^2\over 4}-
{{(y(\tau_a)-y(\tau_b))}^2\over 4\bar T}
\Bigr]
}
\over 
{
{\dps\int}
{\cal D} y
\,\exp\Bigl [- \int_0^T d\tau
{{\dot y}^2\over 4}
\Bigr]
}
}
= {
{{{\rm Det'}({d^2\over {d\tau}^2} - {B_{ab}\over\bar T})}
^{-{D\over 2}}}
\over
{{{\rm Det'}({d^2\over {d\tau}^2})}^{-{D\over 2}}}
}
= {\Bigl(1+{G_{Bab}\over{\bar T}}\Bigr)}^{-{D\over 2}}
\quad.
\label{normchange}
\end{equation}
\no
As usual, the prime denotes the omission of the zero mode from
a determinant.

To summarize, the insertion of a scalar propagator into
a scalar loop can, for fixed values of the proper-time
parameters, be completely taken into account by
changing the path integral normalization, and
replacing $G_B$ by $G_B^{(1)}$.
\no
The vertex operators remain unchanged.

\no
In the case of a photon insertion, 
instead of eq.(~\ref{masslessscalprop}) one has,
in Feynman gauge,

\begin{equation}
-{e^2\over 2}
{{\Gamma \Bigl({D\over 2}-1 \Bigr)}\over {4{\pi}^{D\over 2}}}
{{\dot x(\tau_a)\cdot\dot x(\tau_b)}\over
{\Bigl({[x(\tau_a) - x(\tau_b)]}^2\Bigr)}^{{D\over 2}-1}}\quad 
\label{cci}
\end{equation}
\no
The denominator of the propagator is
treated as above, i.e. exponentiated and absorbed into
the worldline Green's function; the numerator
$\dot x_a\cdot\dot x_b$ remains, and participates in the
Wick contractions.

The whole procedure extends 
without difficulty to the case of $m$ propagator
insertions.
The determinant factor 
eq.(~\ref{normchange}) generalizes
to

\begin{equation}
{\Bigl({\rm det}A^{(m)}\Bigr)}^{D\over 2},
\label{multiloopdet}
\end{equation}
\no
and the two-loop Green's function to

\begin{eqnarray}
G_B^{(m)}(\tau_1,\tau_2) \!\!&=& \!G_B(\tau_1,\tau_2)
\non
&+&
{1\over 2} {\sum_{k,l=1}^m}
\Bigl[G_B(\tau_1,\tau_{a_k})-G_B(\tau_1,\tau_{b_k})\Bigr]
A_{kl}^{(m)}
\Bigl[G_B(\tau_{a_l},\tau_2)-G_B(\tau_{b_l},\tau_2)\Bigr]
           \nonumber\\
\label{nkgk}
\end{eqnarray}
\no
Here $A^{(m)}$ is the
symmetric $m\times m$ -- matrix defined by
\vspace{-1pt}
\begin{eqnarray}
A^{(m)} &=& {\Bigl[\bar T - {C\over 2}\Bigr]}^{-1}
\nonumber\\
\bar T_{kl} &=& \bar T_k \delta_{kl}\nonumber\\
C_{kl} &=& G_B(\tau_{a_k},\tau_{a_l})
- G_B(\tau_{a_k},\tau_{b_l})
- G_B(\tau_{b_k},\tau_{a_l})
+ G_B(\tau_{b_k},\tau_{b_l})\, \nonumber\\
\label{defA}
\end{eqnarray}
\noindent
and
$\bar T_1,\ldots,\bar T_m$ denote the proper-time lengths
of the inserted propagators.

Note that our formulas give the Green's function only between points
on the loop. This suffices, for instance, to calculate single
spinor-loop contributions to photon scattering in QED, i.e. sums
of diagrams as shown in fig. 5, or to the analogous (pseudo-)
scalar scattering diagrams in models with Yukawa couplings
(the generalization to the case of several fermion loops
interconnected by photon propagators requires no new concepts
~\cite{schsch4}). 

In $\phi^3$ -- theory 
or Yang-Mills theory there are, of course, also
contributions to the amplitude with external legs on the
propagator insertions. 
The extension of formula
(\ref{nkgk}) 
involving points located on the insertion 
was given in ~\cite{schsch2} for the
two -- loop case, and obtained for the
general case
by Roland and Sato \cite{rolsat}. 
This knowledge then
is sufficient
to write down
worldline representations for all $\phi^3$ graphs 
which have the topology of a loop with insertions. 
According to graph theory,
for the first few orders of perturbation theory
such a loop, or ``Hamiltonian circuit'', can always
be found
-- all
trivalent graphs with less than 34 vertices do have this property.
$\phi^4$ graphs are reduced to trivalent graphs in the usual way
by introducing an auxiliary field, which can be again represented on the
worldline ~\cite{schsch4}. Even for $\phi^3$ an auxiliary field
must be introduced if one wants to assure that all diagrams are
generated with the appropriate statistical factors ~\cite{schsch4}.
A multiloop generalization for Yang-Mills theory
still waits to be constructed 
(some preliminary steps have been taken in ~\cite{rescsc}).

Roland and Sato also
provided a link back to string theory by 
analyzing the infinite string tension limit of
the Green's function $G_B^{RS(m)}$ of the corresponding 
Riemann surface, and identifying 
$G_B^{(m)}$ with the
leading order term of $G_B^{RS(m)}$
in the ${1\over \alpha'}$ -- expansion:

\be
G_B^{RS(m)}(z_1,z_2)\quad 
{\stackrel{\alpha' \rightarrow 0}{\longrightarrow}}
\quad {1\over \alpha'}
G_B^{(m)}(\tau_1,\tau_2) + {\rm finite}\quad.
\label{Gasympt}
\ee\no
To be precise, $G_B^{(m)}$ 
denotes here not the $G_B^{(m)}$ of
eq.(~\ref{nkgk})
but the equivalent Green's function 
$G_B^{(m)}(\tau_1,\tau_2)-\half
G_B^{(m)}(\tau_1,\tau_1)
-\half G_B^{(m)}(\tau_2,\tau_2)
$.

\vfill\eject
\vskip25pt
{\large\bf 2. Second Lecture: Examples}
\vskip5pt
\vskip5pt
\underline{\bf 2.1) One-loop QED Vacuum Polarization}
\vskip5pt
\no
For a warm-up, let us recalculate the one-loop vacuum polarization
tensors in scalar and spinor quantum electrodynamics 
~\cite{strassler92}.

\vskip5pt\underline{\it A) Scalar QED}
\vskip.3cm

\no
According to eqs. (\ref{scalarpi} -- \ref{planewavebackground}), 
the one-loop two-photon amplitude
in scalar QED can be written as

\bear
\Gamma^{\mu\nu}_{\rm scal}[k_1,k_2]&=&
{(-ie)}^2\Tint\Dx
\int_0^T d\tau_1
\int_0^T d\tau_2\non
&&\times
\dot x^{\mu}(\tau_1)
\e^{ik_1\cdot x(\tau_1)}
\dot x^{\nu}(\tau_2)
{\rm e}^{ik_2\cdot x(\tau_2)}
\freeexp
\label{scalarqed2point}
\ear\no
Separating off the zero mode according to
eqs.(\ref{split}),(\ref{momentumcons}),
one obtains

\bear
\Gamma^{\mu\nu}_{\rm scal}[k_1,k_2]&=&
-{(2\pi )}^D \delta(k_1+k_2)
e^2\Tint\dps
\int_0^T d\tau_1
\int_0^T d\tau_2
\non
&&\times
\int
{\cal D}y\,
\dot y^{\mu}(\tau_1)
\e^{ik_1\cdot y(\tau_1)}
\dot y^{\nu}(\tau_2)
\e^{ik_2\cdot y(\tau_2)}
{\rm e}^{-\int_0^Td\tau {1\over 4}{\dot y}^2}
\; .\non
\label{scalarqed2pointDy}
\ear\no
We use the energy-momentum conservation factor
${(2\pi )}^D \delta(k_1+k_2)$
for setting
$k_1=-k_2=k$, and then omit it.
We now need to Wick-contract the two photon
vertex operators, starting from the
basic rule 

\be
\langle y^{\mu}(\tau_1) y^{\nu}(\tau_2)\rangle
= -g^{\mu\nu}G_B(\tau_1,\tau_2)
\; .
\label{wickscalqed3}
\ee\no
The rules for 
Wick-contracting expressions
involving both
elementary fields and exponentials are
the following:

\begin{enumerate}
\item
Contract fields with each other as usual, and
fields with exponentials according to

\be
\Bigl\langle y^{\mu}(\tau_1)
\e^{ik\cdot y(\tau_2)}\Bigr\rangle
= i\langle y^{\mu}(\tau_1)y^{\nu}(\tau_2)
\rangle k_{\nu}
\e^{ik\cdot y(\tau_2)}
\label{wickfieldexp}
\ee\no
(the field disappears, the exponential stays in the game).

\item
Once all elementary fields have disappeared, 
contraction of the remaining
exponentials yields a universal factor

\be
\Bigl\langle
\e^{ik_1\cdot y_1}
\cdots
\e^{ik_N\cdot y_N}
\Bigr\rangle 
=
\exp\biggl[
-\half
\sum_{i,j=1}^N
k_{i\mu}
\langle
y^{\mu}(\tau_i)
y^{\nu}(\tau_j)
\rangle
k_{j\nu}
\biggr]
\; .
\label{wickNexp}
\ee\no

\end{enumerate}

\no
For the case at hand this produces two terms,

\be
\Bigl\langle
\dot y^{\mu}(\tau_1)
\e^{ik\cdot y(\tau_1)}
\dot y^{\nu}(\tau_2)
\e^{-ik\cdot y(\tau_2)}
\Bigr\rangle
=
\biggl\lbrace
g^{\mu\nu}
\ddot G_{B12}
-k^{\mu}k^{\nu}
\dot G_{B12}^2
\biggr\rbrace
\e^{-k^2G_{B12}}
\; .
\label{scalarqed2pointwick}
\ee\no
Now one could just write out $G_B$ and its derivatives,

\bear
\dot G_B(\tau_1,\tau_2) &=& {\rm sign}(\tau_1 - \tau_2)
- 2 {{(\tau_1 - \tau_2)}\over T}\\
\ddot G_B(\tau_1,\tau_2)
&=& 2 {\delta}(\tau_1 - \tau_2)
- {2\over T}\quad \\
\label{GBderivatives}\nonumber
\ear\no
and calculate the parameter integrals. 
It is useful, though, to first perform a partial integration 
on the term involving $\ddot G_{B12}$ in either
$\tau_1$ or $\tau_2$. The integrand
then turns into

\be
\biggl\lbrace
g^{\mu\nu}k^2
-k^{\mu}k^{\nu}
\biggr\rbrace
\dot G_{B12}^2
\e^{-k^2G_{B12}}
\; .
\label{scalarqed2pointpartint}
\ee\no
Note that this makes the gauge invariance of the
vacuum polarization manifest.
We rescale to the unit circle, 
$\tau_i = Tu_i, i = 1,2$, and use translation
invariance in $\tau$ to fix the zero to 
be at the location of the second vertex operator.
We have then

\begin{equation}
G_B(\tau_1,0)=Tu_1(1-u_1),
\dot G_B(\tau_1,\tau_2)=1-2u_1
.
\label{scaledown}
\end{equation}\no
Taking the free determinant factor eq.(~\ref{freepi})
into account, and
performing the global proper-time integration, one finds  

\bear
\Gamma^{\mu\nu}_{\rm scal}[k]&=&
e^2
\Bigl[k^{\mu}k^{\nu}-
g^{\mu\nu}k^2
\Bigr]
\Tint
{(4\pi T)}^{-{D\over 2}}
T^2
\int_0^1du
{(1-2u)}^2
\e^{-Tu(1-u)k^2}
\non
&=&
{e^2\over {(4\pi )}^{D\over 2}}
\Bigl[k^{\mu}k^{\nu}-g^{\mu\nu}k^2\Bigr]
\Gamma\bigl(2-{D\over 2}\bigl)
\int_0^1du
(1-2u)^2
{\Bigl[
m^2 + u(1-u)k^2
\Bigr]
}^{{D\over 2}-2}
\; .\non
\label{scalarvpresult}
\ear\no
It is easy to verify that this
agrees with the result reached by calculating the
sum of the 
corresponding two field theory diagrams 
in dimensional regularization.

\vskip5pt\underline{\it B) Spinor QED}
\vskip.3cm

\no
For the fermion loop, the path integral for the two-photon amplitude
becomes, in the component formalism,

\vspace{-10pt}
\bear
\Gamma_{\rm spin}^{\mu\nu}[k_1,k_2]&=&
-2{(-ie)}^2\Tint\Dx\Dpsi
\int_0^T d\tau_1
\int_0^T d\tau_2\non
&&
\!\!\!\!\!\!\!\!\!\!\!
\times
\Bigl(
\dot x^{\mu}_1
+2i\psi^{\mu}_1\psi_1\cdot k_1
\Bigr)
\e^{ik_1\cdot x_1}
\Bigl(
\dot x^{\nu}_2
+2i\psi^{\nu}_2\psi_2\cdot k_2
\Bigr)
\e^{ik_2\cdot x_2}
\e^
{-\int_0^Td\tau
\bigl(
{\dot x^2\over 4}
+{\psi\dot\psi\over 2}
\bigr)}
\; .
\non
\label{spinorqed2point}
\ear\no 
The calculation
of ${\cal D}x$ is identical with the scalar QED calculation.
Only the calculation of ${\cal D}\psi$ is new, and amounts to
a single Wick contraction,

\bear
{(2i)}^2
\Bigl\langle
\psi^{\mu}_1\psi_1\cdot k_1
\psi^{\nu}_2\psi_2\cdot k_2
\Bigr\rangle
&=&-
G_{F12}^2
\Bigl[g^{\mu\nu}k^2-k^{\mu}k^{\nu}\Bigr]
\nonumber\\
&=&-\Bigl[g^{\mu\nu}k^2-k^{\mu}k^{\nu}\Bigr]
\; .
\label{spinwick2point}
\ear\no
Adding this to the bosonic result shows that, up to the global
normalization, 
the parameter integral for the spinor loop is obtained
from the one for the scalar loop simply by substituting, in
eq.(~\ref{scalarqed2pointpartint}),

\be
\dot G_{B12}^2 \rightarrow
\dot G_{B12}^2 
- G_{F12}^2 =-{4\over T}G_{B12}
\; .
\label{subs2point}
\ee\no
The complete change thus amounts
to supplying eq.(~\ref{scalarvpresult})
with a global factor of $-2$, and replacing ${(1-2u)}^2$
by $-4u(1-u)$. This leads to

\be
\Gamma_{\rm spin}^{\mu\nu}[k]
=
8{e^2\over {(4\pi )}^{D\over 2}}
\Bigl[k^{\mu}k^{\nu}-g^{\mu\nu}k^2\Bigr]
\Gamma\bigl(2-{D\over 2}\bigl)
\int_0^1du\,
u(1-u)
{\Bigl[
m^2 + u(1-u)k^2
\Bigr]
}^{{D\over 2}-2},
\label{spinorvpresult} 
\ee\no
again in agreement with the result of the
field theory calculation.

The qualitative features of this calculation generalize to the
$N$ -- photon amplitude. It is obvious that the Wick contraction of
$N$ photon vertex operators will lead to an integrand of the form

\be
P_N(\dot G_{Bij},\ddot G_{Bij},G_{Fij})
\exp\biggl[
\sum_{k<l}G_{Bkl}k_k\cdot k_l
\biggr]
\label{polexp}
\ee\no
with some polynomial $P_N$.
What is remarkable is that
the substitution above can be
promoted to the following general
substitution rule. As Bern and Kosower
have shown, it is always possible to remove all second derivatives
$\ddot G_{Bij}$ appearing in the integrand by a suitable chain
of partial integrations. Once this has been done, the integrand
for the spinor loop case can be obtained 
from the one for the scalar loop
by simultaneously
replacing every closed 
cycle of $\dot G_B$'s appearing, say
$\dot G_{Bi_1i_2} 
\dot G_{Bi_2i_3} 
\cdots
\dot G_{Bi_ni_1}$, 
by its ``supersymmetrization'',

\vspace{-10pt}
\begin{equation}
\dot G_{Bi_1i_2} 
\dot G_{Bi_2i_3} 
\cdots
\dot G_{Bi_ni_1}
\rightarrow 
\dot G_{Bi_1i_2} 
\dot G_{Bi_2i_3} 
\cdots
\dot G_{Bi_ni_1}
\nonumber\\
-
G_{Fi_1i_2}
G_{Fi_2i_3}
\cdots
G_{Fi_ni_1}\nonumber\\
\label{subrule}
\end{equation}

\noindent
(up to the global factor of $-2$). 
Note that an expression is considered
a cycle already if it
can be put into cycle form using the
antisymmetry of $\dot G_B$
(e.g. $\dot G_{Bab}\dot G_{Bab}
=-\dot G_{Bab}\dot G_{Bba}$). 
This rule may be understood as a remnant of
worldsheet supersymmetry \cite{berntasi}.

The close connection between the scalar and fermion
loop calculations may appear surprising, as it
appears to have no analogue in standard field theory.
It arises because 
the treatment of fermions in the worldline formalism
(and in string theory) does not correspond to the
usual first order formalism, but to a second order
formalism for fermions. 
The Feynman rules for spinor QED in the second order formalism
(see \cite{morgan} and references therein) are, up to statistics
and degrees of freedom, the ones for scalar QED
with the addition of a third vertex (fig. 8)
\footnote{I thank A.G. Morgan for providing this figure.}.
Note that those rules also apply to the external fermion
case.

\begin{figure}[h]
\begin{center}
{}~\epsfig{file=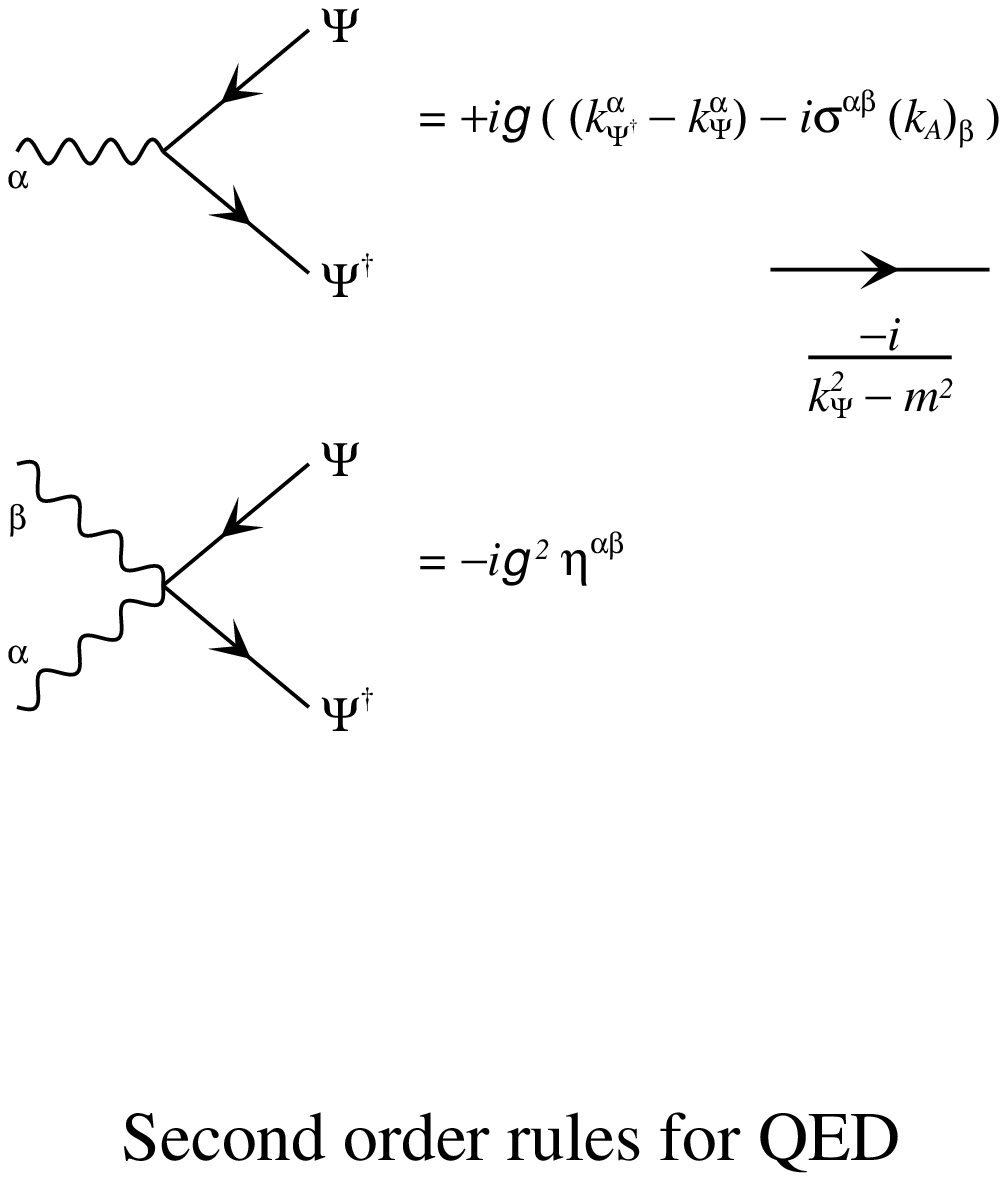,clip=}
\end{center}
\caption[]
{
\label{Fig2ndOrderRules}
The second order Feynman rules for fermion QED. 
They correspond to the ones for scalar QED
with a third vertex added. This vertex involves 
$\sigma^{\mu\nu}={i\over 2}[\gamma^{\mu},\gamma^{\nu}]$
and corresponds to the $\psi^{\mu}F_{\mu\nu}\psi^{\nu}$
-- term in the worldline Lagrangian. For the details 
and for the nonabelian case see
~\cite{morgan}.}
\end{figure}

\vskip5pt
\underline{\bf 2.2) Photon Splitting in QED}
\vskip5pt
\no
Next let us consider an application involving
constant external fields, namely the one-loop
three photon amplitude in a constant magnetic field.
This amplitude gives rise to photon splitting,
a process of potential astrophysical interest
whose first exact calculation was performed
by Adler in 1971 ~\cite{adler71}.

We will again do both the scalar and the spinor QED cases.
This amplitude is finite, so that we can work in $D=4$.
First we have to specialize our formulas 
for the general constant background field
to the pure magnetic field case. We choose the $B$ -- field
along the z -- axis, and introduce matrices
$\bf I_{03}$ and $\bf I_{12}$ projecting on
the $t,z$ -- and $x,y$ -- planes, so that

\begin{equation}
F\equiv
\left(
\begin{array}{*{4}{c}}
0&0&0&0\\
0&0&B&0\\
0&-B&0&0\\
0&0&0&0
\end{array}
\right),
{\bf I_{03}}\equiv
\left(
\begin{array}{*{4}{c}}
1&0&0&0\\
0&0&0&0\\
0&0&0&0\\
0&0&0&1
\end{array}
\right),
{\bf I_{12}}\equiv
\left(
\begin{array}{*{4}{c}}
0&0&0&0\\
0&1&0&0\\
0&0&1&0\\
0&0&0&0
\end{array}
\right).\nonumber\\
\label{defBmatrices}
\end{equation}
\no
We may then rewrite the determinant factors
eqs.(\ref{spindetext})
as

\bear
{\rm det}^{-{1\over 2}}
\biggl[{\sin(eFT)\over {eFT}}
\biggr]&=&
{(eBT)\over{\sinh(eBT)} }
\;\label{scalardetextB},\\ 
{\rm det}^{-{1\over 2}}
\biggl[{\tan(eFT)\over {eFT}}
\biggr]&=&
{(eBT)\over{\tanh(eBT)} }
\; .
\label{spinordetextB}
\ear\no
The Green's functions eq.(\ref{calGBGF}) 
specialize to

\begin{eqnarray}
\bar{\cal G}_{B}(\tau_1,\tau_2) 
&=& G_{B12}{\bf I_{03}}
-{T\over 2}{\Bigl[\cosh(z\dot G_{B12})-\cosh(z)\Bigr]
\over z\sinh(z)}
{\bf I_{12}}\nonumber\\&&
+{T\over{2z}}\biggl({\sinh(z\dot G_{B12})\over\sinh(z)}
-\dot G_{B12}\biggr)i{\hat{\bf F}}\nonumber\\
\dot{\cal G}_{B}(\tau_1,\tau_2)
&=&\dot G_{B12}{\bf I_{03}}+{\sinh(z\dot G_{B12})\over\sinh(z)}
{\bf I_{12}}
-\biggl({\cosh(z\dot G_{B12})\over \sinh(z)}-{1\over z}
\biggr)i{\hat{\bf F}}\nonumber\\
\ddot{\cal G}_{B}(\tau_1,\tau_2)
&=& \ddot G_{B12}{\bf I_{03}}
+2\biggl(\delta_{12}-{z\cosh(z\dot G_{B12})\over T
\sinh(z)}\biggr){\bf I_{12}}
+2{z\sinh(z\dot G_{B12})\over T\sinh(z)}i{\hat{\bf F}}\nonumber\\
{\cal G}_{F}(\tau_1,\tau_2) &=&G_{F12}{\bf I_{03}}
+G_{F12}{\cosh (z\dot G_{B12})\over \cosh (z)}{\bf I_{12}}
-G_{F12}{{\sinh (z\dot G_{B12})}\over{\cosh (z)}}i{\hat{\bf F}}
\nonumber\\
\label{GB(F)pureB}
\end{eqnarray}
\noindent
We have now introduced $z=eBT$, and 
$\hat{\bf F}={F\over B}$.
In writing ${\cal G}_B$ we have already subtracted its
coincidence limit, which is indicated by the ``bar''.
The coincidence limits for $\dot{\cal G}_B$
and ${\cal G}_F$ are also needed,

\begin{eqnarray}
\dot {\cal G}_B(\tau,\tau) &=& -\biggl(\coth(z)
-{1\over z}\biggr)i{\hat{\bf F}}\nonumber\\
{\cal G}_F(\tau,\tau) &=& -\tanh(z)i{\hat{\bf F}}\\
\label{coincidencepureB}\nonumber
\end{eqnarray}
\noindent
To obtain the photon splitting amplitude, we will 
use these correlators for the Wick contraction of
three vertex operators $V_0$ and $V_{1,2}$, representing the
incoming and the two outgoing photons.

\vskip5pt\underline{\it A) Scalar QED}
\vskip.3cm

\noindent
The calculation is
greatly simplified by the peculiar kinematics of this process.
Energy--momentum conservation $k_0+k_1+k_2=0$
forces collinearity of all three four--momenta, so that,
writing $-k_0\equiv k\equiv \omega n$,

\begin{equation}
k_1={\omega_1\over\omega}k,k_2={\omega_2\over\omega}k;\,\,
k^2=k_1^2=k_2^2=k\cdot k_1=k\cdot k_2=k_1\cdot k_2=0.
\label{kconstraints}
\end{equation}

\noindent
Moreover, Adler ~\cite{adler71} was able to show on general grounds
that
there is only one 
non-vanishing polarization case. This is the case
where the magnetic vector 
$\hat {\bf k}\times\hat{\bf\varepsilon_0}$ 
of the incoming photon is
parallel to the plane
containing the external field and the direction of propagation
$\hat {\bf k}$, and those of the
outgoing ones are both perpendicular to this plane. 
An appropriate choice of $\varepsilon_{0,1,2}$ leads to
the further vanishing relations

\begin{equation}
\varepsilon_{1,2}\cdot\varepsilon_0
=\varepsilon_{1,2}\cdot k
=\varepsilon_{1,2}\cdot F=0\quad .
\label{epsconstraints}
\end{equation}

\noindent
This leaves us with the following small
number of nonvanishing Wick contractions:

\begin{equation}
\langle V_0V_1V_2\rangle\!=\! -i\!
\prod_{i=0}^2\int_0^T
\!\!\!d\tau_i
\exp\biggl[{1\over 2}
\!\!\sum_{i,j=0}^2
\bar\omega_i\bar\omega_j
n\bar{\cal G}_{Bij}n\biggr]\,
\varepsilon_1
\ddot{\cal G}_{B12}\varepsilon_2\,
\sum_{i=0}^2
\bar\omega_i\varepsilon_0
\dot{\cal G}_{B0i}n
\label{psscalwick}
\end{equation}

\vskip.3cm
\noindent
For compact notation we have defined
$\bar\omega_0=\omega, \bar\omega_{1,2}=-\omega_{1,2}$.

Performing the Lorentz contractions,
and taking the determinant factor eq.(~\ref{scalardetextB})
into account, one obtains the
following simple parameter integral for this amplitude:

\begin{eqnarray}
C_{\rm scal}[\omega,\omega_1,\omega_2,B] &=&
{m^8\over 8 \omega\omega_1\omega_2}
\int_0^{\infty}\!dT\,T{{\rm e}^{-m^2T}
\over {z}^2{\rm sinh}^2(z)}
\int_0^T\!\!d\tau_1\,d\tau_2\,
\ddot G_{B12}
\non
&&
\!\!\!\!\!\!\!\!\!\!\!\!\!\!\!\!\!\!\!\!\!\!\!\!\!
\times
\biggl[
\sum_{i=0}^2
\bar\omega_i\cosh(z\dot G_{B0i})
\biggr]
\,{\rm exp}
\biggl\lbrace\!\!-{1\over 2}\!\sum_{i,j=0}^2\bar\omega_i
\bar\omega_j\Bigl[G_{Bij} + {T\over 2z}
{{\rm cosh}(z\dot G_{Bij})\over {\rm sinh}(z)}
\Bigr]\biggr\rbrace\nonumber\\
\label{psscalresult}
\end{eqnarray}
\noindent
Translation invariance in $\tau$ has been used to
set the position $\tau_0$ of the
incoming photon equal to $T$. For better comparison
with the literature, we have normalized the amplitude
as in ~\cite{adler71}. 

\vskip5pt\underline{\it B) Spinor QED}
\vskip.3cm
\no
For the fermion loop case, we will now make use of the
superfield formalism. As in the
case without a background field, ${\cal G}_B$
and ${\cal G}_F$ can be combined into a super
propagator,

\be
\hat{\cal G}(\tau_1,\theta_1;\tau_2,\theta_2)
\equiv
{\cal G}_{B 12} + \theta_1\theta_2
{\cal G}_{F12}
\; .
\label{defcalGhat}
\ee\no
This allows us to write the result of the
Wick contraction for the spinor loop in complete
analogy to the scalar loop result 
eq.(\ref{psscalwick}),

\begin{equation}
\langle V_0V_1V_2\rangle\!=\!i\!
\prod_{i=0}^2\int_0^T
\!\!\!d\tau_i\!\!\int\!d\theta_i\,
\exp\biggl[{1\over 2}
\!\!\sum_{i,j=0}^2
\bar\omega_i\bar\omega_j
n\bar{\hat{\cal G}_{ij}}n\biggr]\,
\varepsilon_1
D_1D_2\hat{\cal G}_{12}\varepsilon_2\,
\sum_{i=0}^2
\bar\omega_i\varepsilon_0
D_0\hat{\cal G}_{0i}n
\label{psspinwick}
\end{equation}
\no
Performing the $\theta$ -- integrations
and Lorentz contractions,
we obtain
~\cite{adlsch}:

\begin{eqnarray}
&&C_{\rm spin}
[\omega,\omega_1,\omega_2,B] =
{m^8\over 4 \omega\omega_1\omega_2}
\int_0^{\infty}\!dT\,T{{\rm e}^{-m^2T}
\over {z}^2{\rm sinh}(z)}\nonumber\\
&&\times\int_0^T\!\!d\tau_1\,d\tau_2\,
\,{\rm exp}
\biggl\lbrace\!\!-{1\over 2}\!\sum_{i,j=0}^2\bar\omega_i
\bar\omega_j\Bigl[G_{Bij} + {T\over 2z}
{{\rm cosh}(z\dot G_{Bij})\over {\rm sinh}(z)}
\Bigr]\biggr\rbrace\nonumber\\
&&\times 
\Biggl\lbrace\biggl\lbrack
-\cosh (z)\ddot G_{B12}\!+\!\omega_1\omega_2
\Bigl( {\rm cosh}(z)-\cosh (z\dot G_{B12})\Bigr )\biggr\rbrack
\nonumber\\
&&\times\biggl\lbrack
{\omega\over\sinh(z)\!\cosh(z)}
-\omega_1
{{\rm cosh}(z\dot G_{B01})\over {\rm sinh}(z)}
-\omega_2
{{\rm cosh}(z\dot G_{B02})\over {\rm sinh}(z)}
\biggr\rbrack\nonumber\\
&& +{\omega\omega_1\omega_2
G_{F12}\over \cosh (z)}
\biggl\lbrack
\sinh (z\dot G_{B01})\Bigl (\cosh (z)\!-\!\cosh (z\dot G_{B02}) \Bigr )
- (1 \leftrightarrow 2) 
\biggr\rbrack \Biggl\rbrace.\non
\label{psspinresult}
\end{eqnarray}

\noindent
A numerical analysis of this three-parameter
integral has shown
it to be in agreement with 
other known integral representations of this
amplitude ~\cite{adlsch}. However, the method
of calculation improves in several respects over
standard field theory methods. In particular, it allowed
us to make 
use of the vanishing relations for the kinematic invariants
essentially on line one of the calculation to reduce the number
of terms. 

\vskip5pt
\underline{\bf 2.3) The Two -- Loop QED $\beta$ -- functions}
\vskip5pt
\no
We proceed to the two-loop level, and to a recalculation
of the two-loop $\beta$ -- function coefficients for scalar and
spinor QED. In field theory, the fermion QED calculation requires
consideration of the diagrams shown in fig. 9, while for scalar
QED there are some more diagrams involving the seagull vertex.
Of the possible
counterdiagrams contributions those from
electron wave function and vertex renormalization
cancel on account of the QED Ward identity, however mass
renormalization must be taken into account.
The worldline formalism 
applies only to the calculation of the
bare regularized amplitude; the renormalization has
to be performed in field theory.

\par
\begin{figure}[ht]
\vbox to 4.5cm{\vfill\hbox to 15.8cm{\hfill
\epsffile{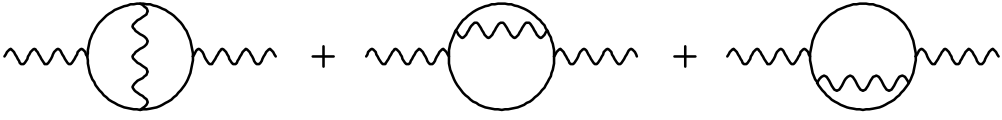}
\hfill}\vfill}\vskip-.4cm
\caption[dum]{Diagrams contributing to the
two-loop QED
vacuum polarization
\hfill}
\label{fig9}
\end{figure}
\par

\noindent
In the following calculation,
which is a variant of the one
presented in ~\cite{schsch3}, 
it will not be necessary to distinguish
between individual graphs.
We will again couple the two-loop path integral to some
constant field background, and write down the 
effective action induced for the background field.
The 
$\beta$ -- function coefficient 
will then be extracted from the 
divergence of the Maxwell
term $\sim F^{\mu\nu}F_{\mu\nu}$, calculated in dimensional regularization.
Of course we are free to choose any background
we wish;
we will therefore impose the condition 
$F^{\mu\nu}F_{\nu\lambda} \sim \delta^{\mu}_{\lambda}$, which
leads to some simplifications.

\vskip5pt\underline{\it A) Scalar QED}
\vskip.3cm
\no
Applying the formalism developed in the first lecture 
to this case, we obtain, for the scalar QED case, immediately
the following parameter integral for the two-loop
effective Lagrangian:

\begin{eqnarray}
{\cal L}^{(2)}_{\rm scal}
[F]&=&
{(4\pi )}^{-D}
\Bigl(-{e^2\over 2}\Bigr)
\int_0^{\infty}{dT\over T}e^{-m^2T}T^{-{D\over 2}} 
\int_0^{\infty}d\bar T 
\int_0^T d\tau_a
\int_0^T d\tau_b
\nonumber\\
&\phantom{=}&\times
{\rm det}^{-{1\over 2}}
\biggl[{\sin(eFT)\over {eFT}}
\biggr]
{\rm det}^{-{1\over 2}}
\biggl[
\bar T 
+{1\over 2}
\Bigl(
\bar{\cal G}_{ab}+
\bar{\cal G}_{ba}
\Bigr)
\biggr]
\langle
\dot y_a\cdot\dot y_b\rangle
\quad .\nonumber\\
\label{Gamma2scal}
\end{eqnarray}
\noindent
Here
$T$ and $\bar T$ denote the scalar
and photon proper-times, and $\tau_{a,b}$ the
endpoints of the photon insertion moving around
the scalar loop. The first determinant factor
is the same as before,
and represents the change of the free
path integral determinant due to the external field;
the second one 
represents its change due to the photon insertion, and generalizes
eq.(~\ref{normchange}) to the
external field case.
The two-loop Green's function eq.(~\ref{G(1)})
generalizes to ~\cite{rescsc}

\begin{equation}
{\cal G}_B^{(1)}(\tau_1,\tau_2)=
{\cal G}_B(\tau_1,\tau_2) + {1\over 2}
{{[{\cal G}_B(\tau_1,\tau_a)-{\cal G}_B(\tau_1,\tau_b)]
[{\cal G}_B(\tau_a,\tau_2)-{\cal G}_B(\tau_b,\tau_2)]}
\over
{{\bar T} +{1\over 2}
\Bigl(
\bar{\cal G}_{ab}
+\bar{\cal G}_{ba}}
\Bigr)}\,
\label{calG(1)} 
\end{equation}
\noindent
We use this Green's function for
Wick-contracting
the ``left over'' numerator of the
photon insertion, which gives

\begin{equation}
\langle
\dot y_a\cdot\dot y_b\rangle
=
{\rm tr}
\biggl[
\ddot{\cal G}_{Bab}+{1\over 2}
{(\dot {\cal G}_{Baa}-\dot {\cal G}_{Bab})
(\dot {\cal G}_{Bab}-\dot {\cal G}_{Bbb})
\over
{\bar T +{1\over 2}
\Bigl(
\bar{\cal G}_{ab}+\bar{\cal G}_{ba}
\Bigr)
}}
\biggr]\; .
\label{wickscal}
\end{equation}
\noindent
Care must be taken again with coincidence limits, as the
derivatives should {\sl not} act on the variables
$\tau_a,\tau_b$ explicitly appearing in the
two-loop Green's function; again the correct rule 
in calculating
$\langle\dot y_a\dot y_b\rangle$ is to first differentiate
eq.(~\ref{calG(1)}) with respect to $\tau_1,\tau_2$,
and set $\tau_1=\tau_a,\tau_2=\tau_b$ afterwards.

Next note that with $F$ chosen as above $\bar{\cal G}_{Bab}
+\bar {\cal G}_{Bba}$ is a Lorentz scalar, so that 

\be
{\rm det}^{-{1\over 2}}
\biggl[
\bar T 
+{1\over 2}
\Bigl(
\bar{\cal G}_{ab}+
\bar{\cal G}_{ba}
\Bigr)
\biggr]
=
{\biggl[
\bar T 
+{1\over 2}
\Bigl(
\bar{\cal G}_{ab}+
\bar{\cal G}_{ba}
\Bigr)
\biggr]}^{-{D\over 2}},
\label{writeoutdet}
\ee\no
and the
$\bar T$ -- integration can be trivially performed.
Note that the term containing $\delta_{ab}$ 
(see eq.(~\ref{calGBGF})) vanishes in dimensional
regularization upon performance of the $\bar T$ -- integral
(it corresponds to a massless
tadpole insertion in field theory). 
After the usual rescaling $\tau_{a,b}=Tu_{a,b}$
one expands the integrand in a Taylor expansion
in $F$, and picks out the coefficient of $\tr(F^2)$.
This gives

\bear
&&{\cal L}^{(2)}_{\rm scal}
[F]={\rm tr}(F^2)
{e^2\over
{(4\pi )}^{D}}
\Bigl(-{e^2\over 2}\Bigr)
\int_0^{\infty}{dT\over T}e^{-m^2T}
T^{4-D}  
\int_0^1 du_a
\int_0^1 du_b
\nonumber\\
&&\!\!\!\!\!\times
\biggl\lbrace
-{1\over 12}{G_{Bab}}^{-{D\over 2}}
+\Bigl[{1\over 2}
-{D-4\over 3(D-2)} -{4\over 3D}
\Bigr]
{G_{Bab}}^{1-{D\over 2}}
+\Bigl[
{28\over 3D}
-{8\over D-2}
\Bigr]
{G_{Bab}}^{2-{D\over 2}}
\biggr\rbrace.\non
\label{scalparint}
\ear
\no
The scalar proper-time integral gives the usual
global $\Gamma(4-D)$, and the  
remaining parameter
integral produces Euler Beta -- functions:

\be
\int_0^1 du_a
\int_0^1 du_b\,
{G_{Bab}}^{n-{D\over 2}}
=
\int_0^1du_a
{[u_a(1-u_a)]}^{n-{D\over 2}}
=
B\Bigl(
n+1-{D\over 2},
n+1-{D\over 2}
\Bigr).
\label{eulerbeta}
\ee\no
Expanding the result in $\epsilon=D-4$,
the ${1\over \epsilon^2}$ -- terms cancel, as expected.
The ${1\over\epsilon}$ -- term becomes

\be
{\cal L}^{(2)}_{\rm scal}[F]
\sim
{1\over2\epsilon}
{\alpha^2\over
{(4\pi)}^2}
F^{\mu\nu}F_{\mu\nu}
\; .
\label{scalmaxdiv}
\ee\no
So far this is a calculation of the bare regularized amplitude.
It must still be supplemented with a contribution from
one-loop mass renormalization, which
involves only one -- loop quantities (see \cite{schsch3}),

\be
\Delta {\cal L}^{(2)}_{\rm scal}[F]
= \delta {m^{(1)}}^2
{\partial\over \partial m^2}
{\cal L}^{(1)}_{\rm scal}[F]\nonumber\\
\sim 
{1\over {2 \epsilon}}
{\alpha^2\over{(4\pi)}^2}
F_{\mu\nu}F^{\mu\nu}
\; .
\label{scalarmassrenorm}
\ee\no
Adding up both contributions

\be
{\cal L}^{(2)}_{\rm scal}[F]+
\Delta {\cal L}^{(2)}_{\rm scal}[F]
\sim
{1\over\epsilon}
{\alpha^2\over{(4\pi)}^2}
F_{\mu\nu}F^{\mu\nu}
\label{scalmaxdivtot}
\ee\no
and extracting the $\beta$ -- function
coefficient in the usual way
(see e.g. ~\cite{ditreueffact}), one finds
the known result ~\cite{bialynicka65},

\begin{equation}
\beta^{(2)}_{\rm scal}(\alpha )= 
{{\alpha}^3\over{2{\pi}^2}}
\quad .
\label{scal2loopcoeff}
\end{equation}
\no

\vfill\eject
\vskip5pt\underline{\it B) Spinor QED}
\vskip.3cm

\no
For the spinor loop in the superfield formalism
one obtains again parameter
integrals formally analogous to
eqs.(~\ref{Gamma2scal}),(~\ref{wickscal}):

\begin{eqnarray}
{\cal L}^{(2)}_{\rm spin}
[F]&=&
(-2)
{(4\pi )}^{-D}
\Bigl(-{e^2\over 2}\Bigr)
\int_0^{\infty}{dT\over T^{1+{D\over 2}}}
e^{-m^2T}
\int_0^{\infty}d\bar T 
\int_0^T d\tau_a d\tau_b
\int d\theta_a d\theta_b
\nonumber\\
&\phantom{=}&\times
{\rm det}^{-{1\over 2}}
\biggl[{\tan(eFT)\over {eFT}}
\biggr]
{\rm det}^{-{1\over 2}}
\biggl[
\bar T 
+{1\over 2}
\Bigl(
{\bar{\hat{\cal G}}}_{ab}
+{\bar{\hat{\cal G}}}_{ba}
\Bigr)
\biggr]
\langle
-D_aY_a\cdot D_bY_b\rangle
\quad ,
\nonumber\\
\label{Gamma2spin}
\end{eqnarray}
\no

\be
\langle
-D_aY_a\cdot D_b Y_b\rangle
=
{\rm tr}
\biggl[
D_aD_b\hat{\cal G}_{Bab}+{1\over 2}
{D_a( \hat{\cal G}_{Baa}- \hat{\cal G}_{Bab})
D_b( \hat{\cal G}_{Bab}- \hat{\cal G}_{Bbb})
\over
{\bar T +{1\over 2}
\Bigl(
{\bar{\hat{\cal G}}}_{ab}
+{\bar{\hat{\cal G}}}_{ba}
\Bigr)
}}
\biggr]\; ,
\label{wickspin}
\ee\no
The further calculation also parallels the scalar case.
The only point to be mentioned is that it is permissible and
convenient to perform the $\bar T$ -- integration {\sl before}
the $\theta$ -- integrals. The equivalent of 
eq.(~\ref{scalparint}) becomes

\bear
&&{\cal L}^{(2)}_{\rm spin}
[F]={\rm tr}(F^2)
{e^4\over
{(4\pi )}^{D}}
\int_0^{\infty}{dT\over T}e^{-m^2T}
T^{4-D}  
\int_0^1 du_a
\int_0^1 du_b
\nonumber\\
&&\!\!\!\!\!\!\!\!\!\!
\times
\biggl\lbrace
{D\over 6}{G_{Bab}}^{-{D\over 2}}
+\Bigl[
{D\over 6}-4
+{8\over 3D}
+{8\over 3(D-2)}
\Bigr]
{G_{Bab}}^{1-{D\over 2}}
+\Bigl[
{28\over 3D}
-{8\over D-2}
\Bigr]
{G_{Bab}}^{2-{D\over 2}}
\biggr\rbrace.\non
\label{spinparint}
\ear
\no
This yields 

\be
{\cal L}^{(2)}_{\rm spin}[F] \sim 
-{3\over \epsilon}{\alpha^2\over{(4\pi)}^{2}}
F_{\mu\nu}F^{\mu\nu}
+ O({\epsilon}^0) \qquad .
\label{spinmaxdiv}
\ee\no
Mass renormalization contributes a
~\cite{schsch3}
\be
\Delta {\cal L}^{(2)}_{\rm spin}[F]
= \delta m^{(1)}
{\partial\over \partial m}
{\cal L}^{(1)}_{\rm spin}[F]
\sim 
{4\over \epsilon}
{\alpha^2\over {(4\pi)}^2}
F_{\mu\nu}F^{\mu\nu}
\; ,
\label{spinormassrenorm}
\ee\no
yielding a total of 

\be
{\cal L}^{(2)}_{\rm spin}[F] + 
\Delta {\cal L}^{(2)}_{\rm spin}[F]
\sim
{1\over {\epsilon}}
{\alpha^2\over {(4\pi)}^2}
F_{\mu\nu}F^{\mu\nu}
\label{spinmaxdivtot}
\ee\no
The $\beta$ -- function coefficient becomes

\begin{equation}
\beta^{(2)}_{\rm spin}(\alpha ) 
=  {{\alpha}^3\over{2{\pi}^2}}
\quad ,
\label{spin2loopcoeff}
\end{equation}
\no
which is the classical Jost-Luttinger
result ~\cite{joslut}.

I have chosen to show the most straightforward, though not the
most efficient versions of this calculation. 
Both in the scalar and the spinor QED case the 2-loop
$\beta$ -- function calculation can be further
trivialized by suitable partial integrations
in $\tau_a$. The absence of a subdivergence
for the Maxwell term 
then becomes manifest, 
regularization is only needed for
the trivial global $T$ -- integration,
and all $\tau_a,\tau_b$ -- dependence
disappears from the integrand before the $\tau_a,\tau_b$
-- integration. This is discussed elsewhere
~\cite{schsch3,rescsc,schsch4}.

\vskip5pt
\underline{\bf 2.4) The Three -- Loop Scalar Master Integral}
\vskip5pt
\no
Finally, let 
us have a look at
the simplest example of a three -- loop
parameter integral calculation in this formalism.
This is the one where
the integrand consists just of the 
bosonic three-loop determinant factor, eq.(~\ref{multiloopdet})
with $m=2$.
In dimensional regularization, it reads

\be
{\Gamma}^{(3)}_{\rm vac}(D)=
\Tint T^{6-{3\over 2}D}I(D),
\label{defI3}
\ee\no

\begin{equation}
I(D)=\int_0^{\infty}\!\!\! d\hat T_1\, d\hat T_2
\int_0^1 \!\!\! da\, db\, dc\, dd\,
{\biggl[(\hat T_1+G_{Bab})(\hat T_2+G_{Bcd})-{C^2\over 4}\biggr]}
^{-{D\over 2}}
\label{defID}
\end{equation}

\noindent
Here $\hat T_{1,2}={T_{1,2}\over T}$ denote the proper-time lengths of the
two inserted propagators in units of $T$, and
$C\equiv G_{Bac}-G_{Bad}-G_{Bbc}+G_{Bbd}$.

This is the most basic integral 
appearing if one
applies this formalism to the calculation of three -- loop
renormalization group functions in abelian
field theories, e.g. in QED ~\cite{flscsc2}
or the Yukawa Model.
As far as the 
quenched QED $\beta$ -- function is concerned, 
the general parameter integral appearing
at the three - loop level
still has a very similar structure:
It has the same universal denominator, 
possibly with a different power, and a numerator,
which is again expressible in terms of the functions
$G_{Bab}$, $G_{Bcd}$, $C$ and their derivatives.

In writing eq.({~\ref{defI3}) we have already rescaled to
the unit circle, and separated off the
electron proper-time integral.
This integral decouples, and just yields
an overall factor of

\begin{displaymath} 
\int_0^{\infty}{dT\over T}{\rm e}^{-m^2T}T^{6-{3\over 2}D}
=\Gamma(6-{3\over2}D) \, m^{3D-12}
\sim -{2\over 3\epsilon}\quad .
\end{displaymath}

\noindent
The nontrivial integrations are 
$\int_0^1da\,db\,dc\,dd
\equiv\int_{abcd}$, 
representing the four propagator
end points 
moving around the 
loop (fig.10). 

\begin{figure}[h]
\begin{center}
\begin{picture}(0,0)%
\epsfig{file=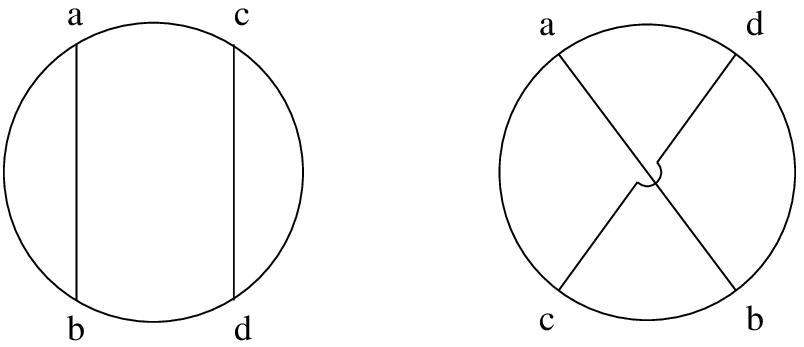}%
\end{picture}%
\setlength{\unitlength}{0.00087500in}%
\begingroup\makeatletter\ifx\SetFigFont\undefined
\def\x#1#2#3#4#5#6#7\relax{\def\x{#1#2#3#4#5#6}}%
\expandafter\x\fmtname xxxxxx\relax \def\y{splain}%
\ifx\x\y   
\gdef\SetFigFont#1#2#3{%
  \ifnum #1<17\tiny\else \ifnum #1<20\small\else
  \ifnum #1<24\normalsize\else \ifnum #1<29\large\else
  \ifnum #1<34\Large\else \ifnum #1<41\LARGE\else
     \huge\fi\fi\fi\fi\fi\fi
  \csname #3\endcsname}%
\else
\gdef\SetFigFont#1#2#3{\begingroup
  \count@#1\relax \ifnum 25<\count@\count@25\fi
  \def\x{\endgroup\@setsize\SetFigFont{#2pt}}%
  \expandafter\x
    \csname \romannumeral\the\count@ pt\expandafter\endcsname
    \csname @\romannumeral\the\count@ pt\endcsname
  \csname #3\endcsname}%
\fi
\fi\endgroup
\begin{picture}(3634,1536)(201,-850)
\end{picture}
\caption{\label{fig10} \hspace{.6cm} 
$\mbox{10a}$ \hspace{4.5cm} $\mbox{10b}$ \hspace{2cm}}
\end{center}
\end{figure}

\noindent
This fourfold
integral decomposes into
24 ordered sectors, of which 16 
constitute the planar (P)
(fig. 10a) and 8 the nonplanar (NP)
sector (fig. 10b). Due to the symmetry
properties of the integrand, all sectors
of the same topology give an equal
contribution.  
The integrand has a trivial invariance
under the operator
${\partial\over\partial a}+
{\partial\over\partial b}+
{\partial\over\partial c}+
{\partial\over\partial d}
$,
which just shifts the location of the
zero on the loop.

As a first step in the calculation 
of $I(D)$, it is useful to
add and subtract the same integral
with $C=0$ and rewrite 

\be
I(D)=I_{\rm sing}(D)+I_{\rm reg}(D),
\label{splitI(D)}
\ee\no
\begin{eqnarray}
&& I_{\rm sing}(D)\!=\!\!\!
\int_0^{\infty}\!\!\!
d\hat T_1\,d\hat T_2\!\!\!\!\!
\int\limits_{\hspace{4mm} abcd}
{\Bigl[(\hat T_1+G_{Bab})(\hat T_2+G_{Bcd})\Bigr]}
^{-{D\over 2}}
\label{defIsing}
\end{eqnarray}

\noindent
$I_{\rm sing}(D)$ 
factorizes into two 
identical three-parameter
integrals, which are elementary:

\begin{displaymath}
I_{\rm sing}(D)=
{\biggl\lbrace
\int_0^{\infty}
dT\int_0^1 du
{\Bigl[
T+u(1-u)
\Bigr]}^{-{D\over 2}}
\biggr\rbrace}^2=
\biggl[{2B(2-{D\over 2},2-{D\over 2})\over D-2}\biggr]^2
\end{displaymath}

\noindent
The point of this split is that the remainder
$I_{\rm reg}(D)$ is finite. To see this,
set $D=4$, 
expand the original integrand in 
${C^2\over G_{Bab}G_{Bcd}}$, and note
that for all terms but the first one
the zeroes of $G_{Bab}$ ($G_{Bcd}$)
at $a\sim b$ ($c\sim d$) are offset by
zeroes of $C^2$.
Only the $1\over\epsilon$ -- pole  
is required in
renormalization group function calculations,
so that one can set $D=4$ 
for the calculation of $I_{\rm reg}(D)$.
The integrations over $\hat T_1,\hat T_2$ are then elementary,
and we are left with

\begin{eqnarray}
I_{\rm reg}(4)\hskip-.3cm&=&\hskip-.7cm
\int\limits_{\hspace{4mm} abcd}
\biggl[-{4\over C^2}{\rm ln}
\Bigl(1\!-\!{C^2\over 4G_{Bab}G_{Bcd}}\Bigr)
\!-\!{1\over G_{Bab}G_{Bcd}}\biggr]
\label{Ireg(4)}
\end{eqnarray}

\noindent
For the calculation of this integral, observe
the following simple behaviour of the function $C$
under the operation
$D_{ab}\equiv {\partial\over\partial\tau_a}
+{\partial\over\partial\tau_b}$:

\begin{equation}
D_{ab}C= \pm 2\chi_{NP},\;
D_{ab}^2C= 2
\Bigl(
\delta_{ac}-\delta_{ad}-\delta_{bc}+\delta_{bd}
\Bigr)
\; ,
\label{DC}
\end{equation}

\noindent
where 
$\chi_{NP}$ denotes the 
characteristic function
of the nonplanar sector.
From these identities and the symmetry properties 
one can easily derive the
following projection identities, which effectively
integrate out the variable $C$:

\begin{eqnarray}
\int_P f(C,G_{Bab},G_{Bcd})
&=&4\int_0^1da\int_0^adc(a-c)
f\Bigl(-2c(1-a),a-a^2,c-c^2\Bigr),\non
\int_{NP}f(C,G_{Bab},G_{Bcd})&=&
\!\!\!
-4\int_0^1da\int_0^adc\int_0^{-2c(1-a)}
dCf\Bigl(C,a-a^2,c-c^2\Bigr).
\non
\label{project}
\end{eqnarray}
\noindent
Here $f$ is an arbitrary function in the variables
$G,G_{Bab},G_{Bcd}$, and 
$\int_0^C dC f$ denotes the integral of this
function in the variable $C$, with the other variables fixed.
The integrals on the left hand side are restricted to the
sectors indicated.
For $f$ the integrand of our formula eq.~(\ref{Ireg(4)}),
we have

\begin{eqnarray}
\int_0^C dC f & = &-{C\over G_{Bab}G_{Bcd}}+
{4\over C} \ln \Bigl( 1 -{C^2\over4 G_{Bab}G_{Bcd}} \Bigr) 
\nonumber\\
& & \hspace{1cm} + {4\over\sqrt{G_{Bab}G_{Bcd}}
} \, \mbox{arctanh} \Bigl( {1\over2} {C\over
\sqrt{G_{Bab}G_{Bcd}}} \Bigr)
\label{integrand}
\end{eqnarray}

\noindent
Inserted in the second equation of (~\ref{project}) this
leaves us with three two--parameter integrals, of
which the first one is elementary.
Applying the substitution

\begin{displaymath}
y={c(1-a)\over a(1-c)}  
\end{displaymath}
\noindent
to the second integral, and

\begin{displaymath}
y^2={c(1-a)\over a(1-c)}
\end{displaymath}
\no
to the third integral, those are
transformed into known standard integrals,
tabulated for instance in ~\cite{devduk}.
The result is

\begin{displaymath}
\int_{NP}f=12\zeta(3) -8\zeta(2) \quad .
\end{displaymath}

\noindent
The calculation in the planar sector is elementary, and I just
give the result,

\begin{displaymath}
\int_Pf = 4\zeta(2)-4 \quad .
\end{displaymath}

\noindent
Putting the pieces together, we have,
up to terms of order $O(\epsilon^0)$,

\begin{displaymath}
{\Gamma}^{(3)}_{\rm vac}(D)=
m^{3D-12}
\Gamma(6-{3\over 2}D)\,
\biggl\lbrace
\Bigl[{2B(2-{D\over 2},2-{D\over 2})\over D-2}\Bigr]^2
+12\zeta(3)-4\zeta(2)-4
\biggr\rbrace
.
\end{displaymath}
\vskip-.2cm
\noindent

\vskip15pt
{\bf Acknowledgements}

\no
I thank M. Jezabek and the organizers of the school for 
their hospitality and the opportunity to deliver these
lectures. I am also grateful to M. Reuter and
M. G. Schmidt 
for discussions, 
and to D. Fliegner for computer support.

\newpage

\vskip2cm
{\bf Particle Path Integrals in Quantum Field Theory}

\end{document}